\documentclass[a4paper,11pt]{article}

\usepackage{comment}
\usepackage{physics}
\usepackage{cancel}
\usepackage{mathtools}
\usepackage{aas_macros}
\usepackage{float}
\pdfoutput=1 

\usepackage{jcappub} 

\usepackage[T1]{fontenc} 

\usepackage{tikz}
\usetikzlibrary{shapes.geometric, arrows}
\tikzstyle{box0} = [rectangle, rounded corners, minimum width=3cm, minimum height=1cm,text centered, draw=black, fill=gray!50]
\tikzstyle{box1} = [rectangle, rounded corners, minimum width=4cm, minimum height=.5cm,text centered, draw=black, fill=blue!40]
\tikzstyle{box2} = [rectangle, rounded corners, minimum width=4cm, minimum height=.5cm,text centered, draw=black, fill=purple!50]
\tikzstyle{box3} = [rectangle, rounded corners, minimum width=4cm, minimum height=.5cm,text centered, draw=black, fill=orange!50]
\tikzstyle{box4} = [rectangle, rounded corners, minimum width=4cm, minimum height=.5cm,text centered, draw=black, fill=yellow!70]
\tikzstyle{arrow} = [thick,->,>=stealth]

\title{\boldmath The Hierarchical Cosmic Web and Assembly Bias}

\author[1,2,3]{J. M. Coloma-Nadal,}
\author[1,2]{F.-S. Kitaura}
\author[1,2]{J. E. García-Farieta,}
\author[4,5,1,2]{F. Sinigaglia,}
\author[1,2]{G. Favole}
\author[6]{and D. Forero Sánchez}

\affiliation[1]{Instituto de Astrof\'isica de Canarias, \\ Calle Via L\'actea s/n, E-38205, La  Laguna, Tenerife, Spain}
\affiliation[2]{Departamento  de  Astrof\'isica, Universidad de La Laguna, \\  E-38206, La Laguna, Tenerife, Spain}
\affiliation[3]{Institute of Space Sciences (ICE, CSIC), Campus UAB, Carrer de Can Magrans, s/n, 08193 Barcelona, Spain}
\affiliation[4]{Département d’Astronomie, Université de Genève, Chemin Pegasi 51, CH-1290 Versoix, Switzerland}
\affiliation[5]{Institut für Astrophysik, Universität Zürich, Winterthurerstrasse 190, CH-8057 Zürich, Switzerland}
\affiliation[6]{Institute of Physics, Laboratory of Astrophysics, École Polytechnique Fédérale de Lausanne (EPFL), Observatoire de Sauverny, CH-1290 Versoix, Switzerland}

\emailAdd{ jmcoloma@ice.csic.es, fkitaura@iac.es}

\abstract{Accurate modeling of galaxy distributions is paramount for cosmological analysis using galaxy redshift surveys. However, this endeavor is often hindered by the computational complexity of resolving the dark matter halos that host these galaxies. To address this challenge, we propose the development of effective assembly bias models down to small scales, i.e., going beyond the local density dependence capturing non-local cosmic evolution. We introduce a hierarchical cosmic web classification that indirectly captures up to third-order long- and short-range non-local bias terms. This classification system also enables us to maintain positive definite parametric bias expansions. Specifically, we subdivide the traditional cosmic web classification, which is based on the eigenvalues of the tidal field tensor, with an additional classification based on the Hessian matrix of the negative density contrast. We obtain the large-scale dark matter field on a mesh with $\sim3.9\,h^{-1}$  Mpc cell side resolution through Augmented Lagrangian Perturbation Theory. To assess the effectiveness of our model, we conduct tests using a reference halo catalogue extracted from the UNIT project simulation, which was run within a cubical volume of 1 $h^{-1}$ Gpc side. The resulting mock halo catalogs, generated through our approach, exhibit a high level of accuracy in terms of the one-, two- and three-point statistics. They reproduce the reference power-spectrum within better than 2 percent accuracy up to wavenumbers $k\sim0.8\,h$ Mpc$^{-1}$ and provide accurate bispectra within the scales that are crucial for cosmological analysis. This effective bias approach provides a forward model appropriate for field-level cosmological inference and holds significant potential for facilitating cosmological analysis of galaxy redshift surveys, particularly in the context of projects such as DESI, EUCLID, and LSST.}

\begin{document}
\maketitle
\flushbottom

\section{Introduction}
\label{sec:intro}

\textcolor{black}{The main goal of modern cosmology is to unveil the nature of the main components of the Universe, such as dark matter and dark energy. For this purpose, galaxy surveys like DESI ~\citep{levi2013desi} or EUCLID ~\citep{laureijs2011euclid} are aimed at mapping the large-scale of the Universe observing millions of galaxies over huge volumes. The central purpose of this endeavor is to measure the influence of dark energy on the expansion of the Universe as well as the effect of dark matter in the evolution of structures. To achieve this objectives, these surveys demand a robust analysis of observational data by means of a strong theoretical framework and galaxy mock catalogues, which are essential for estimating covariance matrices and study systematic effects.}

Generating mock catalogues require the faithful recreation of the cosmic volumes observed in actual galaxy surveys, along with the replication of clustering observables. Traditionally, simulating structure formation has relied on $N$-body codes. However, when working with the immense volumes involved in contemporary surveys, this is not the most suitable scenario. As a practical alternative, approximate gravity solvers come into play. Among these, some use Lagrangian Perturbation Theory (LPT) \citep{Bernardeau_2002} to compute the dark matter density field. Noteworthy examples of these approaches include in chronological order:  \texttt{PEAKPATCH} \citep{Bond_1996,Bond_1996b,Bond_1996c,Stein_2019},  \texttt{PINOCCHIO} \citep{Monaco_2002,Monaco_2013}, \texttt{PThalos} \citep{Scoccimarro_2002,Manera_2013,Manera_2015}, \texttt{PATCHY} \citep{Kitaura_2014,Kitaura_2015,Kitaura_2016,Vakili_2017}, \texttt{EZmocks} \citep{Chuang_2014,Zhao_2021}, \texttt{HALOGEN} \citep{Avila:2014nia}, \texttt{ICE-COLA} \cite{Izard_2016}, and \texttt{BAM} \citep{Balaguera_2018,Balaguera_2020,Kitaura_2022,Balaguera_2023}, each offering distinct methodologies for generating galaxy mock catalogues  (see comparison projects, \citep{Chuang_2015,Lippich_2019,Blot_2019,Colavincenzo_2019}). 

These methods can be classified into two categories: the ones trying to reproduce the halo distribution based on analytical and numerical approximations (\texttt{PEAKPATCH}, \texttt{PINOCCHIO}, later versions of \texttt{PThalos}, \texttt{HALOGEN}, \texttt{ICE-COLA}); and those trying to reproduce the distribution of large-scale structure tracers from detailed reference calculations with effective field-level bias models (e.g., the earliest \texttt{PThalos} version, \texttt{PATCHY}, \texttt{EZmocks}, and \texttt{BAM}).

The latter category offers substantial savings in computational resources, as demonstrated in a comparative study by \citep{Blot_2019}. Moreover, it enables a detailed dissection of contributions from individual bias components, as elucidated in works by \citep{Kitaura_2015,Kitaura_2022}. Importantly, this methodology can be likewise extended to field-level inference studies (see, e.g.,\citep{Kitaura_2021,Ata_2015}).

In this study, our approach to modelling structure formation hinges on augmented Lagrangian Perturbation Theory (ALPT) ~\citep{Kitaura_2013}, an advancement over linear LPT. ALPT refines the displacement field by decomposing it into long- and short-range components, drawing from second-order LPT (2LPT) and the spherical collapse approximation, respectively. This nuanced treatment allows for a more precise characterisation of the evolving cosmic structures within a one-step gravity solver. We could, however, rely on more accurate and  expensive approaches (e.g., eALPT\citep{Kitaura_2024}, \texttt{FASTPM} \citep{Feng_2016}, and \texttt{COLA} \citep{Tassev_2013}).  The findings of the present study will yield to improved results when using more sophisticated gravity solvers, as demonstrated in \citep{Vakili_2017}.

The advantage of \texttt{PATCHY} over \texttt{EZmocks} and \texttt{BAM} is its explicit, clear analytic bias description, i.e., the description of the intricate relationship between the dark matter density field and its tracers such as halos, galaxies,  galaxy clusters, etc. 
However, \texttt{PATCHY} faced challenges in accurately modelling the higher-order statistics of low-mass tracers, particularly halos with masses above approximately $10^8 M_\odot\,h^{-1}$ in a cubical volume of 400 $h^{-1}$ Mpc side, despite incorporating second-order non-local bias effects \citep{Pellejero_2020}.

The study by \citep{Kitaura_2022}, using high mass halos  (halos with $\gtrsim10^{12} M_\odot\,h^{-1}$  in a 1,5 $h^{-1}$ Gpc side cubical volume), showed that a proper modelling of the higher order statistics requires going to at least third-order non-local bias, and that this can be done through the cosmic web classification. These findings were exploited in the context of effective baryonic physics bias mapping and Lyman-$\alpha$ forest modelling (see \citep{Sinigaglia_2021,Sinigaglia_2022,Sinigaglia_2024}).

This paper extends previous \texttt{PATCHY} versions with a novel hierarchical cosmic web classification, indirectly accounting for both long- and short-range non-local bias at least up to third order. This bias description turns out to be crucial in acquiring a high fidelity forward model, as tested on a $N$-body based halo catalogue with a mass cut of $\gtrsim 10^{11} M_\odot\,h^{-1}$ in a 1 $h^{-1}$ Gpc side cubical volume with variance suppression \citep{Chuang_2019MNRAS}.

Our method offers the advantage of streamlining the parametric bias model, consolidating it into a maximum of six parameters that encompass both local deterministic and stochastic bias components.  Moreover, the indirect incorporation of non-local bias terms through the hierarchical cosmic web classification enhances the model's comprehensiveness and efficiency. Our approach avoids the explicit inclusion of non-local bias terms, a practice that can lead to the introduction of negative densities when truncated to a specific order. Instead, we segment a large-scale structure catalogue into well-defined subsets, akin to dividing a halo catalogue into mass bins. This segmentation allows us to treat each subset as a distinct catalogue, simplifying the application of a straightforward local bias model to accurately reproduce its characteristics.
In this way, we manage to have a solid description of the large scale bias on large scales as stated by renormalised perturbation theory \cite{McDonald_2009,2006PhRvD..73f3519C}, and extend it towards small scales with short-range non-local bias terms. The bias model presented in this study could potentially be combined with 
 effective field theory based approaches \cite{Baumann_2012,Carrasco_2012,Pajer_2013,Porto_2014,Angulo_2015,Senatore_2015,Senatore_2015b,Vlah_2015} within field-level frameworks (see, e.g., \cite{Schmittfull_2019}).

While our approach concentrates solely on number counts within a real-space mesh, it is worth noting that a companion paper introduces a sub-grid model designed to allocate positions and velocities within the cells (see the CosmoMIA method \citep{Forero_2024}).  The bias method presented in this paper together with CosmoMIA are being applied to a variety of large-scale structure tracers, such as Luminous Red Galaxies (LRGs), Emission Line Galaxies (ELGs), quasars (QSOs) and Bright Galaxies (BGs), and will be presented in forthcoming publications. 

The paper is structured as follows. In section \ref{sec:method},  
we briefly describe the structure formation theory we use to evolve the density field. Subsequently, we present  the bias of dark matter tracers through perturbation theory (PT) and define our analytic model. At the end of this section we hierarchically classify the cosmic web. In section \ref{sec:calibration}, we present and statistically analyse the test on a $N$-body reference catalog. Finally, we discuss the results in section \ref{summary}.

\section{Theory}
\label{theory}

In this section, we provide an overview of the analytical expression of the displacement field necessary to calculate the density field. Then we recap the theoretical background in modelling the bias of large scale structure tracers.

\subsection{Structure formation}

On the large scales, we rely on  Lagrangian perturbation theory (LPT) to simulate  structure formation \citep{Bernardeau_2002}.  This is an approximate solution to evolve particles from the primordial density fluctuations until the desired epoch (redshift). In this theory, the action of gravity across cosmic evolution is encoded in the displacement field $\mathbf{\Psi}(\mathbf{q},z)$, which relates initial Lagrangian positions $\mathbf{q}$ to final Eulerian comoving ones  $\mathbf{x}(z)$ through the following mapping relation
\begin{equation}
    \mathbf{x}(z) = \mathbf{q} + \mathbf{\Psi}(\mathbf{q}, z).
    \label{eq:ALPT}
\end{equation}
However, it is well known in the literature that LPT approaches suffer from artificial shell-crossing, as the particles do not interact with each other \citep{Buchert_1994,Bouchet_1995,Catelan_1995,Crocce_2006}. This problem can be mitigated with Augmented LPT (ALPT) \citep{Kitaura_2013}, a  combination of LPT with the spherical collapse (SC) model \citep{Bernardeau_1994, Mohayaee_2006, neyrinck_2013}, which models virialisation towards small scales. 
In particular, we rely on second-order LPT (2LPT) \citep{Buchert_1994, Bouchet_1995}, an improvement of the linear approximation with non-local tidal field corrections.
The ALPT displacement field is then written as sum of two components
\begin{equation}
    \mathbf{\Psi}(\mathbf{q}, z) = \underbrace{\overbrace{\mathcal{K}(\mathbf{q}, r_s)}^{\text{long-range kernel}}\circ\mathbf{\Psi}_{\mathrm{2LPT}}(\mathbf{q}, z)}_\text{long-range interaction} + \underbrace{\overbrace{\left(1 -\mathcal{K}(\mathbf{q}, r_s)\right)}^{\text{short-range kernel}}\circ\mathbf{\Psi}_{\mathrm{SC}}(\mathbf{q}, z)}_\text{short-range interaction}
    \label{eq:displacementfield}
\end{equation}
where the  radius $r_s$ smoothly separates both regimes \textcolor{black}{and takes the value 6 $h^{-1}$ Mpc in this work, as the typical uncertainty in the LPT solution is of the order of 3 $h^{-1}$ Mpc \citep[][]{Monaco_2013}. In this way, the LPT solution is gradually substituted by the spherical collapse solution, mitigating artificial shell-crossing.}
The long-range component is described by the linear growth factor $D$ and the first- and second-order gravitational potential terms $\phi^{(1)}$, $\phi^{(2)}$:
\begin{equation}
    \mathbf{\Psi}_{\mathrm{2LPT}} = -D\nabla\phi^{(1)}+D_2\nabla\phi^{(2)}.
\end{equation}
The short-range component reads
\begin{equation}
    \mathbf{\Psi}_{\rm SC}=\nabla\nabla^{-2}\left[3\left(\left(1-\frac{2}{3}D\delta^{(1)} \right)^{1/2}-1 \right) \right].
\end{equation}
For more details, we refer the reader to \citep{Kitaura_2013}. The smooth dark matter density field $\delta^{\rm ALPT}$ is finally obtained by applying Cloud-In-Cell (CIC) interpolation to the final positions $\mathbf{x}(z)$ including tetrahedra-tesselation \citep{Abel_2012}. The ALPT method  presents a better cross-correlation with $N$-body simulations than applying separately either LPT solutions until third order or the spherical collapse model alone \citep{Kitaura_2013,Neyrinck_2016}. Furthermore, compared to $N$-body solutions, it is considerably more efficient, \footnote{See table 1 in \citep{Blot_2019}: comparison between \texttt{ICE-COLA} and \texttt{PATCHY} to reproduce the Minerva simulations \citep{Minerva_2016}: 33 hours CPU time with 340 Gb memory allocation vs 0.2 hours and 15 Gb, respectively.} 
while reproducing the power on the very large scales at different redshifts. 

Our studies have shown that the usage of improved gravity solvers, \textcolor{black}{such as eALPT \citep{Kitaura_2024}, an extension of ALPT,} makes only sense when the resolution of the particle-mesh is higher than about $2.5\,h^{-1}$ Mpc, \textcolor{black}{since otherwise the calculation of the gravitational potential towards small scales is too inaccurate no matter which gravity solver is used (see also a study comparing ALPT with particle-mesh gravity solvers at resolutions of $2.6\,h^{-1}$ Mpc side cells, where moderate gain is reported \citep[][]{Vakili_2017})}. Since we want to explore at first, the maximal efficiency of the method, we explore in this paper only resolutions of $\sim3.9 \,h^{-1}$ Mpc. Conveniently, we opt for a resolution that is a multiple of two ($2^4$) lower than the original dark matter field {\color{black} in each dimension}, which serves as our reference simulation from the UNIT project \citep{Chuang_2019MNRAS}. We leave an exploration of our framework with higher resolutions based on eALPT for future work.

\subsection{The bias of dark matter tracers}

Galaxies, galaxy clusters and halos are biased tracers of the dark matter distribution (for review, see \citep{Desjacques_2018}). The relations between different tracers  and the underlying dark matter field, named \textit{bias relations}, are well known to be non-linear, non-local (hence, also scale-dependent)  and stochastic  (see \citep{Kitaura_2014} and references therein). Bias relations also depend on the history evolution of the Universe since the tracer formed, the so-called assembly bias (see, e.g., \citep{Gao_2005, Croton_2007, Dalal_2008}). Assembly bias is also studied through the dependence of the tracer on the cosmic web (see, e.g., \cite{Tojeiro_2017,Yang_2017}).

On the large scales, where density fluctuations are small, perturbation theory (PT) can be used, following \citep{Fry_1993}, to expand the relation between \textcolor{black}{the tracer number density}  $\delta_{\rm tr}$ and the dark matter density $\delta$,
\begin{equation}
 \delta_{\rm tr}(\mathbf{x}) = f(\delta(\mathbf{x})) = \sum_{i=0}^{\infty}\frac{b_i}{i!}\delta^i(\mathbf{x}),
 \label{eq:taylor_expansion}
\end{equation}
where $b_i$ are the unknown bias parameters. This expansion, however, presents only a dependence on the local density. In order to allow for history dependence and long-range non-locality, one can also include a dependency on the peculiar velocity through $\partial_i v_j$ and on the gravitational potential through the tidal field tensor $\mathcal{T}_{ij}\equiv\partial_i\partial_j\phi(\mathbf{x})$ \citep{McDonald_2009, Kitaura_2022}. 
In this work, we extend local bias models  including long- and short-range non-local bias components through a cosmic web classification. The short-range non-local bias is based on the second-order derivatives of the density field, $\Gamma_{ij}\equiv\partial_i\partial_j\delta(\mathbf{x})$. The dependencies we investigate can be summarised as follows
\begin{equation}
    \delta_{\rm tr} \curvearrowleft P\left(\delta_{\rm tr} | \nabla^2\phi(\mathbf{x}),       \partial_i\partial_j\phi(\mathbf{x} ), \partial_i\partial_j\delta(\mathbf{x} ), \epsilon \right) \equiv P\left(\delta_{\rm tr} |\delta, \mathcal{T}, \Gamma, \epsilon\right),
    \label{eq:dependencies}
\end{equation}
where $\epsilon$ describes stochastic uncertainty arising from the fact that the tracer distribution is a discrete sample of the underlying dark matter distribution. By convenience, following \citep{McDonald_2009}, the tidal field tensor can be rewritten by removing its trace
\begin{equation}
    s_{ij}(\mathbf{x})\equiv \mathcal{T}_{ij}-\frac{1}{3}\delta_{ij}^K\delta(\mathbf{x})\,.
    \label{eq:s-tensor}
\end{equation}
Furthermore, the cosmological principle (homogeneity and isotropy) constrains the bias parameters to be constant scalar quantities. Then, one should construct scalars from $s_{ij}$. The first order scalars $s_{ii}$ are zero by definition and the second and third order terms can be constructed by the tensor contractions $s^2\equiv s_{ij}s_{ji}$ and  $s^3\equiv s_{ij}s_{jk}s_{ki}$, respectively. 
Analogously, we can define a traceless tensor for the short-range non-local bias term,
\begin{equation}
    \mu_{ij}(\mathbf{x})\equiv \Gamma_{ij}-\frac{1}{3}\delta_{ij}^K\nabla^2\delta(\mathbf{x}),
    \label{mu-tensor}
\end{equation}
and its scalars $\mu^2\equiv \mu_{ij}\mu_{ji}$ and $\mu^3\equiv \mu_{ij}\mu_{jk}\mu_{ki}$.

The resulting Taylor expansion of the bias model, in Eulerian coordinates $\mathbf{x}$, has now local ($B_{\rm L}$), non-local long-range ($B^{\rm LR}_{\rm NL}$), non-local short-range ($B^{\rm SR}_{\rm NL}$), velocity ($B^{v}_{\rm NL}$) and stochastic ($B^{\epsilon}$) bias contributions. Following \citep{Kitaura_2022} and \citep{McDonald_2009}, the expansion can be written as
\begin{equation}
\footnotesize
\begin{split}
\delta_{\rm tr}(\mathbf{x})= &B(\delta_{\rm tr}|\delta)=\overbrace{B_{\rm L}(\delta)}^{\text{local}}+\overbrace{\overbrace{B_{\rm NL}^{\rm LR}(\delta)}^{\text{long-range}}+\overbrace{B_{\rm NL}^{\rm SR}(\delta)}^{\text{short-range}}+\overbrace{B_{\rm NL}^{v}(\delta)}^{\theta\neq\delta}}^{\text{non-local}}+\overbrace{\overbrace{B^{\epsilon}(\delta)}^{\text{stochastic noise}}}^{\text{local\;\&\;non-local}}
\\=&\underbrace{
\underbrace{
\overbrace{
c_\delta\delta(\mathbf{x})
}^{\mathrm{local}}
}_{\mathrm{first\; order}} 
+
\underbrace{
\overbrace{
\frac{1}{2} c_{\delta^2}\left( \delta^2(\mathbf{x})-\langle\delta^2\rangle\right)
}^{\mathrm{local}} 
+
\overbrace{
\frac{1}{2} c_{s^2} \left(s^2(\mathbf{x})-\frac{2}{3}\langle\delta^2\rangle\right)
}^{\mathrm{non-local\;long-range}}
}_{\mathrm{second\; order}}  
+ 
\underbrace{
\overbrace{
\frac{1}{3!} c_{\delta^3} \delta^3(\mathbf{x})
}^{\mathrm{local}} 
+
\overbrace{
\frac{1}{2}c_{\delta s^2} \delta(\mathbf{x}) s^2(\mathbf{x}) +
\frac{1}{3!} c_{s^3} s^3(\mathbf{x})
}^{\mathrm{non-local\;long-range}}
}_{\mathrm{third\; order}}}_{\mathrm{curl-free\;\&\;\theta=\delta\;terms}:\,B_{\rm L}|^{(1\to3)}+B^{\rm LR}_{\rm NL}|^{(2\to3)}}
\\
& \hspace{-1.1cm}+\overbrace{\underbrace{\underbrace{\frac{1}{2}c_{\mu^2}\left(\mu^2(\mathbf{x}) - \langle\mu^2(\mathbf{x}) \rangle \right)}_{\mathrm{second\;order}} + \underbrace{\frac{1}{2}c_{\delta\mu^2}\delta(\vec x)\mu^2(\mathbf{x}) + \frac{1}{3!}c_{\mu^3}\mu^3(\mathbf{x})}_{\mathrm{third\;order}}}_{\mathrm{curl-free\;\&\;\theta=\delta\;terms}:\,B^{\rm SR}_{\rm NL}|^{(2\to3)}}}^{\mathrm{non-local\;short-range}} 
+
\underbrace{
\underbrace{\overbrace{\underbrace{
\mathcal{O}(F_{v}(\mathbf{x})|^{(3)})
}_{\mathrm{third\; order+}}}^{\mathrm{non-local}}}_{\mathrm{\theta\neq\delta\,terms}}}
_{{\cal O}(B^{v}_{\rm NL}|^{(3)})}
+ 
\underbrace{
\underbrace{\overbrace{\underbrace{
\mathcal{O}(F_{\delta,\mu,\mathcal T}(\mathbf{x})|^{(4)})}_{\mathrm{fourth\; order+}}}^{\mathrm{local\;\&\;non-local}}}_{\mathrm{curl-free\,\theta=\delta\,terms}
}}
_{
{\cal O}(\delta_{\rm tr}|^{(4)})}
+ 
\underbrace{
\underbrace{\underbrace{\overbrace{F_{\epsilon}(\epsilon(\mathbf{x}))}^{\mathrm{local\;\&\;non-local}}}_{\mathrm{first\;order+}}}_{\mathrm{noise\;terms}
}}
_{
{\cal O}(B^{\epsilon}(\delta)|^{(1)})}\,,
\end{split}
\label{eq:bias_expansion}
\end{equation}
where the $c$s are the bias factors and the superscripts $(1\to3)$ denote terms from first to third order.

 In the large-scale limit of $k\to0$, the observable, renormalised, large scale bias $b_\delta$ as the sum of bias-like terms according to \cite{McDonald_2009} is given by
\begin{equation}
\label{eq:blarge}
b_{\delta} = c_{\delta} + \left( \frac{34}{21} c_{\delta^2} + \frac{68}{63} c_{s^2} + \frac{1}{2} c_{\delta^3} + \frac{1}{3} c_{\delta s^2} - \frac{16}{63} c_{s t} \right) \sigma^2 + \frac{1}{2} c_{\delta\epsilon^2} \sigma^2_\epsilon\,,
\end{equation}
where $\sigma^2$ is the variance of the dark matter field at the considered resolution, $\sigma^2_\epsilon$ is the corresponding variance of the noise, and $c_{st}$ relates to a mixed non-local bias term including velocity contributions for $\theta\neq\delta$.

\section{Method}
\label{sec:method}

Equation \eqref{eq:bias_expansion} encapsulates the principal bias terms of the tracer density field $\delta_{\rm tr}$ within a Taylor expansion. However, employing truncated versions of an infinite expansion at a practical level is non-trivial and carries inherent risks.

As depicted in \eqref{eq:blarge}, accurate reproduction of large-scale bias necessitates accounting for at least third-order bias contributions. Yet, this does not offer guidance on transitioning from large to small scales to achieve a comprehensive field-level bias description.

As elucidated in \citep{Kitaura_2022}, the contributions of velocity bias primarily impact small scales and can generally be disregarded for the majority of baryon acoustic oscillations (BAO) and redshift-space distortions (RSD) studies. Consequently, we defer an investigation into extending our bias model to incorporate terms where $\theta \neq \delta$ for future endeavors. Instead, we focus on terms where information from velocity shear is inherently encoded in the tidal field tensor. 

Here, we adopt the approach proposed in the aforementioned study and introduce, for the first time in the context of parametric bias models of discrete tracer distributions, a cosmic web classification to indirectly model non-local bias terms. We extend this methodology to similarly model short-range non-local bias contributions for the first time, as elaborated below.

This approach has several advantages, as it permits us to ensure positive definite tracer density fields, and directly or indirectly include all bias terms ($B_{\rm L}$, $B_{\rm NL}^{\rm LR}$, $B_{\rm NL}^{\rm SR}$, $B_{\rm NL}^{v}$ and $B^{\epsilon}$). Our method consists on applying a deterministic (section \ref{:deterministic}) and stochastic (section \ref{stochastic}) bias model to each of the different regions of the cosmic web which is hierarchically classified in terms of the gravitational potential $\phi$ and the density field $\delta$ (section \ref{sec:non-local bias}). 

The deterministic bias, consists firstly on a power law \citep{Cen_1993,delatorre_2013,Kitaura_2014} and secondly on a threshold bias \citep{Kaiser_1984,Kitaura_2014,Neyrinck_2014}, describing the non-linear behaviour and allowing to suppress low density regions where not enough mass is available to form halos. We extend this picture with a suppression of density peaks which some galaxies tend to avoid as a result of halo exclusion (see, e.g., \cite{Garcia_2019}). 

Finally, the connection between the expected number counts and the discrete realisation of large scale structure tracers is established through a likelihood, accounting for  stochastic biasing \cite{Dekel_1999,Sheth_1999}.
As advanced by \citep{Peebles_1980}, a non-vanishing correlation function  introduces deviations from Poissonity \cite{Somerville_2001,Casas_Miranda_2002}. This affects the statistics at scales below the mesh resolution and needs to be modelled with non-Poisson likelihoods  \cite{Saslaw_1984,Sheth_1995,Kitaura_2014,Pellejero_2020}

\subsection{Local deterministic bias}
\label{:deterministic}
Following \citep{Kitaura_2014}, we can generally write the expected number counts of a certain tracer $\langle \rho_{\rm tr} \rangle$ in a given volume element d$V$ as function of the underlying dark matter density $\delta$,
\begin{equation}
    \langle \rho_{\rm tr}\rangle_{\mathrm{d} V}= f_{\rm tr}(1+B(\delta_{\rm tr}|\delta)),
\end{equation}
where $B(\delta_{\rm tr}|\delta)$ is the deterministic bias relation and $f_{\rm tr}$ is the normalisation factor which controls the number density and can be described as
\begin{equation}
    f_{\rm tr}=\frac{n_{\rm tr}}{\langle 1+B(\delta_{\rm tr}|\delta)\rangle_V},
\end{equation}
i.e., by requiring the tracer density field to have the number density of the reference sample $\langle\langle\rho_{\rm tr}\rangle_{\mathrm{d}V}\rangle_V =n_{\rm tr}$.

In particular, we define the bias relation as the product of a power-law, describing the non-linear behaviour, an exponential damping for high-density regions ($\epsilon>0$) and a low-density threshold ($\epsilon'<0$),
\begin{equation}
    1+B_{\rm L}(\delta)=\underbrace{(1+\delta)^{\alpha}}_{\text{power-law bias}}\times \underbrace{\exp\left(-\left(\frac{1+\delta}{\rho_{\rm \epsilon}}\right)^{\epsilon}\right)}_{\text{high-density threshold bias}}\times \underbrace{\exp\left(-\left(\frac{1+\delta}{\rho^{\prime}_{\rm \epsilon}}\right)^{\epsilon^{\prime}}\right)}_{\text{low-density threshold bias}}\,,
    \label{eq:deterministic_bias}
\end{equation}
where we have extended the model presented in \citep{Kitaura_2014} to also account for high density suppression.

\subsection{Stochastic bias}
\label{stochastic}

The distribution of halos on a dark matter density field defined on a mesh, characterised by a lower resolution than that needed to resolve the halos, leads to a stochastic process. This can be described by the probability $P(N_i|\lambda_i)$ of finding a given number of halos $N_i$ in a given cell $i$ with volume $dV$. Therefore,  $\lambda_i=\langle\rho_{\rm tr}\rangle_{\rm d V}\times \mathrm{d}V$ is the average number of particles in a cell. The Poisson distribution, $P(N_i|\lambda_i)=\frac{\lambda_i^{N_i}}{N_i!}\exp(-\lambda_i)$, can account for the discrete nature of galaxies \citep{wild_2005, de_la_Torre_2013}. However, it only  approximately holds in intermediate density regions. In general,  we have over-dispersion. Non-Poissonian distributions, such as the Negative Binomial (NB), where introduced to account for field-level bias modelling \cite{Kitaura_2014,Ata_2015,Schmittfull_2019}:
\begin{equation}
    P(N_i|\lambda_i, \beta)=\underbrace{\frac{\lambda_i^{N_i}}{N_i!}\exp{(-\lambda_i)}}_{\text{Poisson distribution}}\times\underbrace{\frac{\Gamma(\beta+N_i)}{\Gamma(\beta)(\beta+\lambda_i)^{N_i}}\frac{\exp(\lambda_i)}{(1+\lambda_i/\beta)^{\beta}}}_{\text{deviation from Poissonity}}
    \label{eq:NB}\,.
\end{equation}
The Poisson distribution is recovered in the limit $\beta\rightarrow \infty$ for which the deviation tends to one \citep{Vakili_2017}. One can also model a density dependent $\beta$-parameter to account for arbitrary variance expressions (see \cite{Pellejero_2020}).
In this work, we model the stochastic bias by drawing samples of the deterministic bias model \eqref{eq:deterministic_bias} from the NB distribution.

\subsection{Non-local deterministic bias: The hierarchical cosmic web classification}
\label{sec:non-local bias}

Numerical simulations have shown that the gravitational growth of the initial density perturbations leads to a characteristic pattern of structures, dubbed cosmic web \citep{Bond_1996nat,van_de_Weygaert_2008}, very similar to the one from observations of the large-scale distribution of galaxies (see, e.g., \citep{Gott_2005}). Most of the Universe is composed of low-density regions (voids) surrounded by filaments of matter generating a web-pattern. Inspired by the formation of cosmic sheet-like {\it pancake} structures \citep{Zeldovich_1970},  \citep{Hahn_2007}  proposed a theoretical criterion to classify the different regions of the cosmic web in terms of the particle motion inside the gravitational potential of a dark matter halo, $\ddot{x}=-\nabla\phi$. The center of mass of a halo is an extremum, i.e. $\nabla\phi=0$, and hence, there, the equation of motion can be linearised. At this points, the particle motion is completely determined by the tidal field tensor $\mathcal{T}_{ij}\equiv\partial_i\partial_j\phi$ and, more concretely, by its eigenvalues $\lambda_1>\lambda_2>\lambda_3$. Analogously to the \textit{Zel'dovich pancake}, we consider regions with three positive eigenvalues as those where gravitational collapse cause matter inflows. On the other hand, we consider unstable orbits (expanding regions) when the three eigenvalues are negative. Attending to this argument, one can define four regions in the cosmic web
\begin{itemize}
    \item  Knots: $\lambda_1$, $\lambda_3$, $\lambda_3>$ $\lambda_{\rm th}$
    \item  Filaments: $\lambda_1$, $\lambda_2>$ $\lambda_{ \rm th}$ $\&$ $\lambda_3<$$\lambda_{\rm th}$
    \item  Sheets: $\lambda_1>$  $\lambda_{\rm th}$ $\&$ $\lambda_2$, $\lambda_3<$ $\lambda_{ \rm th}$
    \item  Voids: $\lambda_1$, $\lambda_3$, $\lambda_3<$$\lambda_{\rm th}$
\end{itemize}

\begin{figure}[ht!]
    \centering
    \includegraphics[scale=0.2]{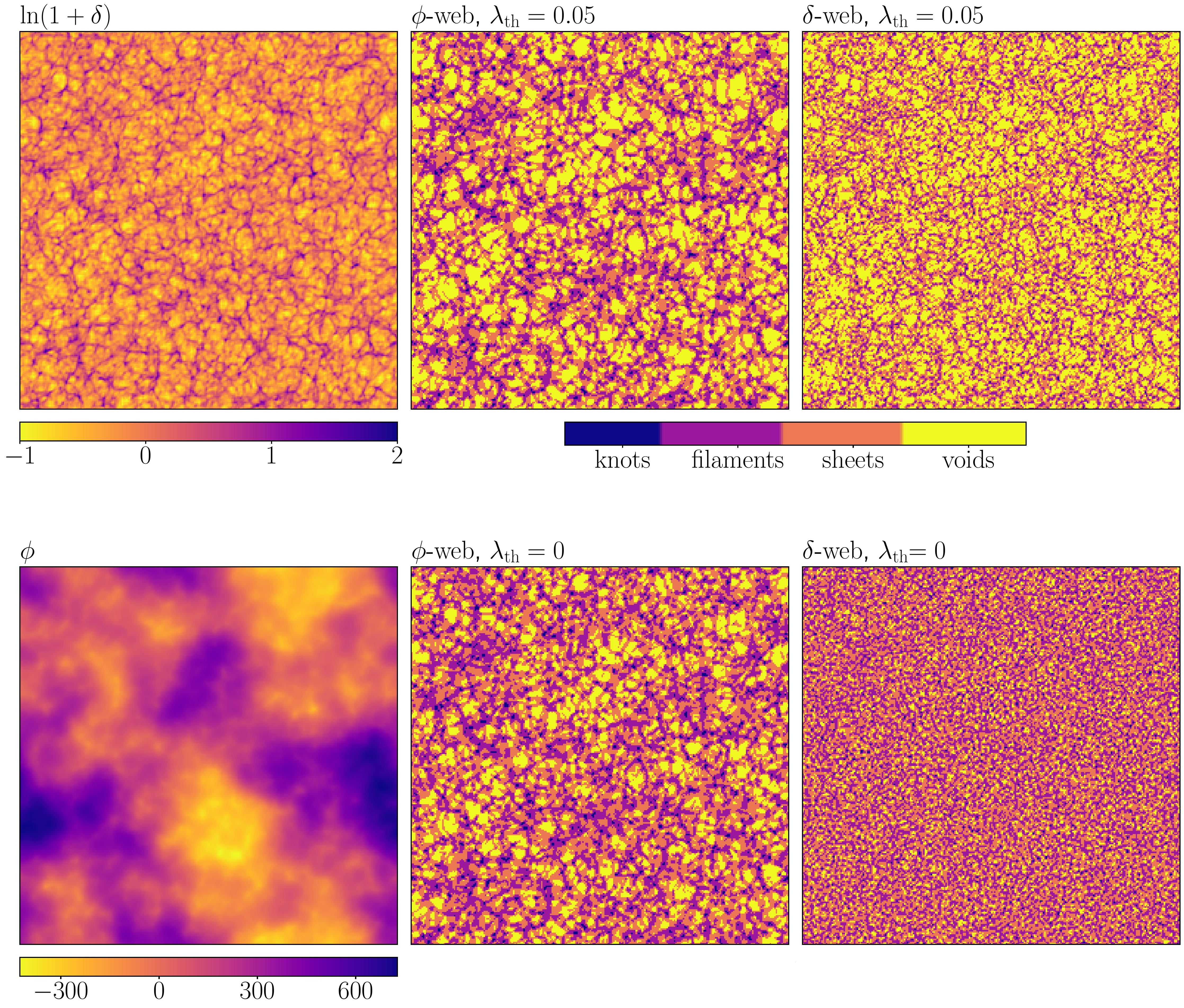}
    \caption{{\color{black} Upper panels: ALPT Dark matter density field and the chosen large-scale and small-scale cosmic web fields for this study with $\lambda_{\rm th}=0.05$ for both cases.} Slices of 20 $h^{-1}$ Mpc thick and 1000 $h^{-1}$ Mpc side at redshift $z$ = 1.032 are shown for the logarithmic density field ({\color{black}upper} left panel), the $\phi$-web ({\color{black}upper} middle panel) and the $\delta$-web ({\color{black}upper} right panel). The darker the colour, the higher the density. {\color{black} Lower panels: Same as upper panels, but showing the gravitational potential (lower left panel), and the large-scale (lower middle panel) and small-scale (lower right panel) cosmic web fields, with $\lambda_{\rm th}=0$ for both cases.}}
    \label{fig:COSMICWEB}
\end{figure}
The \textcolor{black}{dimensionless} threshold eigenvalue $\lambda_{\rm th}$\footnote{\color{black}The gravitational potential $\phi$ is obtained from solving the dimensionless Poisson equation $\nabla^2\phi=\delta$. In consequence, the eigenvalues of the tidal field tensor and $\lambda_{\rm th}$ are dimensionless too. Note that the first invariant of the tidal field tensor yields the density contrast: $\delta =\lambda_1+\lambda_2+\lambda_3$ (see equations \ref{eq:inv} and \ref{eq:deltas}).}, allowing for $\lambda_{\rm th} \neq 0$, was introduced by \citep{Forero_2009} to reduce the dependence on the smoothing scale, mesh resolution and mass assignment scheme. This classification can be generalised to the full density field rather than limit to the centre of the halos. As is essentially defined by the gravitational potential, we dubbed this classification $\phi$-web (see figure \ref{fig:COSMICWEB}). It has been used in simulations both to study the properties of dark matter halos, galaxies and intergalactic gas \citep{Hahn_2007, Yang_2017, Martizzi_2019, Sinigaglia_2021} and to generate halos and galaxy mock catalogues  \citep{Zhao_2015, Balaguera_2018, Pellejero_2020}. Furthermore, it is used in observational analysis \citep{Lee_2016, Krolewski_2017, Horowitz_2019}.  

The gravitational potential is a key ingredient in both the bias model \eqref{eq:bias_expansion} and the cosmic web classification. The distribution of halos can be associated with various regions of the cosmic web by considering the invariants of the tidal field tensor. These invariants, in turn,  directly  describe the non-local bias terms within halo perturbation theory (for details see \citep{Kitaura_2022} and references therein), and are:
\begin{equation}
\label{eq:inv}
\begin{split}
    &I_1 = \lambda_1+\lambda_2+\lambda_3, \\
    &I_2 = \lambda_1\lambda_2+\lambda_1\lambda_3+\lambda_2\lambda_3, \\
    &I_3 = \lambda_1\lambda_2\lambda_3. \\
\end{split}
\end{equation}

Consequently, the invariants can be directly linked to the different regions of the cosmic web (for simplicity, and without loss of generality,  for $\lambda_{\rm th}=0$):
\begin{itemize}
    \item    Knots: $I_3>0\:\&\: I_2>0\:\&\:I_1>\lambda_1$
    \item  Filaments: $\left(I_3<0 \:\&\:I_2<0\right) \;||\; \left( I_3<0\:\&\:I_2>0 \:\&\: \lambda_3<I_1<\lambda_3-\lambda_2\lambda_3/\lambda_1 \right)$
    \item   Sheets: $\left(I_3>0 \:\&\:I_2<0 \right)\;||\:\left( I_3<0 \:\&\:I_2>0 \:\&\:\lambda_1-\lambda_2\lambda_3/\lambda_1<I_1<\lambda_1\right)$
    \item  Voids: $I_3<0\:\&\:I_2>0\:\&\: I_1<\lambda_1$
\end{itemize}

It emerges that both the long-range non-local bias  \eqref{eq:bias_expansion} and the cosmic web classification are determined by the eigenvalues of the tidal field. 
In particular, the invariants can be related to the different bias components  in halo perturbation theory (see \cite{Kitaura_2022})
\begin{equation}
\label{eq:deltas}
\begin{split}
      &\delta=I_1, \\
    &s^2 = \frac{2}{3}I_1^2-2I_2, \\
     &s^3 = -I_1I_2+3I_3+\frac{2}{9}I_1^3.
     \end{split}
\end{equation}
The cosmic web classification based on the gravitational potential (the $\phi$-web) provides information on $I_1\sim\delta$, $I_2$ and $I_3$, and thus, on $\delta$, $s^2$, and $s^3$.  Thus, we propose applying the deterministic \eqref{eq:deterministic_bias} and stochastic \eqref{eq:NB} bias model to the different $\phi$-web regions to indirectly model long-range non-local bias. 
Additionally, since the short-range non-local bias \eqref{eq:bias_expansion} is determined by the density field tensor, $\Gamma_{ij}=\partial_i\partial_j\delta$, we suggest sub-classifying the $\phi$-web regions based on the eigenvalues of the density field tensor.
Figure \ref{fig:COSMICWEB} illustrates {\color{black} in the upper panels} the density field (left) alongside the $\phi$-web classification (middle) and the $\delta$-web classification (right). 
A visual examination of figure \ref{fig:COSMICWEB} reveals the correlation between both classifications and the density field, with the $\delta$-web providing information on small scales. {\color{black} The lower panels in figure \ref{fig:COSMICWEB} show how the gravitational potential traces larger scales (left), and how the $\phi$-web classification (middle) is quite stable against moderate variations in $\lambda_{\rm th}$, while the $\delta$-web classification is very noisy when the $\lambda_{\rm th}$ is set to zero.}

\begin{figure}[ht!]
    \centering
    \includegraphics[scale=0.22]{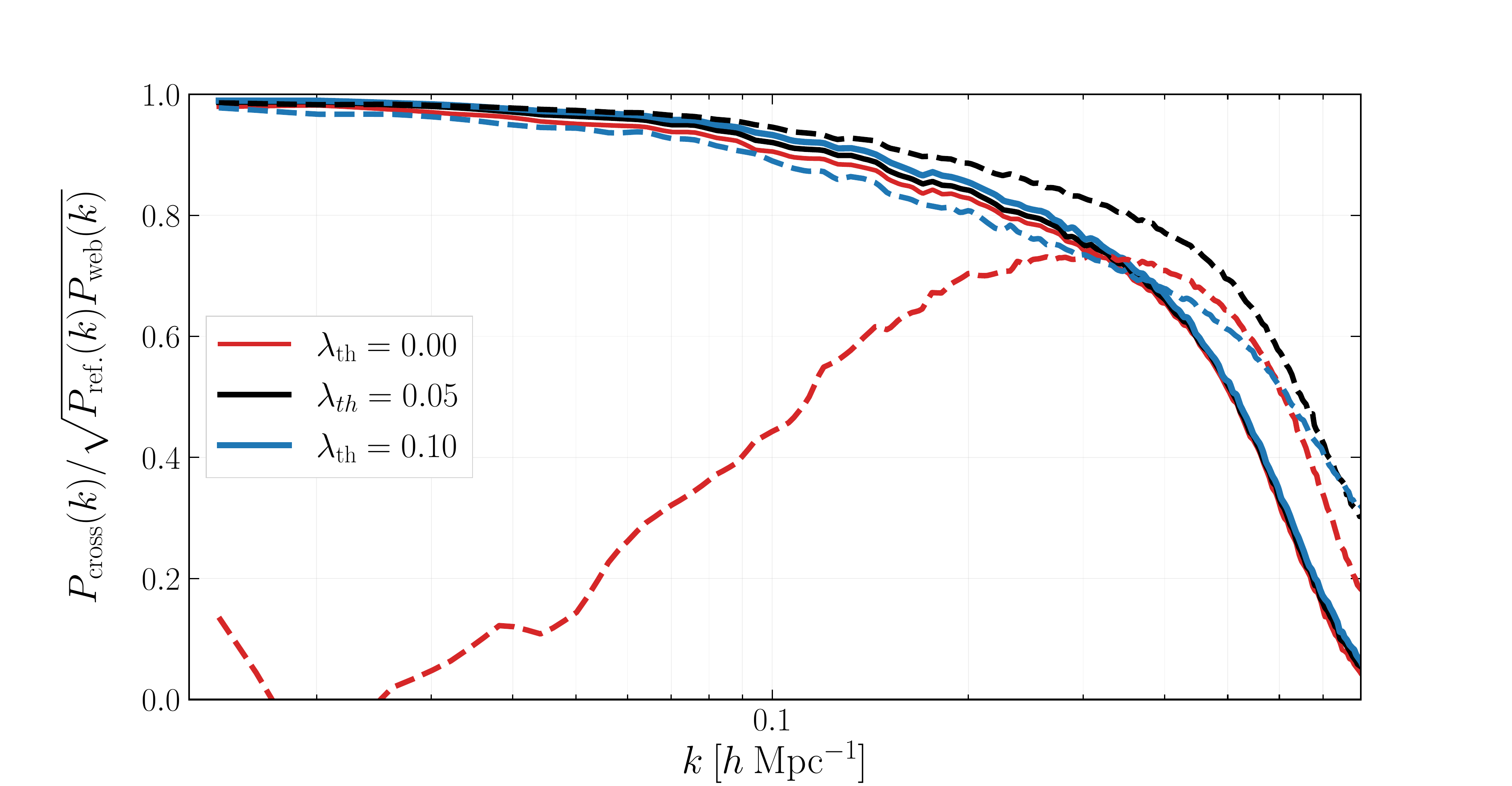}
    \caption{Cross power-spectra of the cosmic web density fields $\delta_{\phi-\text{web      }}$ (solid lines) and $\delta_{\delta-\text{web}}$ (dashed lines) with the halo reference catalogue for different threshold values $\lambda_{\rm th}$.  }
    \label{fig:CROSSPK}
\end{figure}

\textcolor{black}{We have employed an information theory-based approach to represent the ALPT density field using only four distinct values, corresponding to different cosmic web regions, in order to study the cross-correlation with the halo reference catalogue and obtaining a criterion to determine the most suitable value for the $\lambda_{\rm th}$ parameter. Following the eigenvalues, we assign the numbers 4, 3, 2 and 1 to, respectively, knots, filaments, sheets and voids for both $\phi-web$ and $\delta-web$ classification. An inverted order of numbers yields anticorrelated fields.}
The specific values themselves are not critical; what is important is ensuring that the difference between subsequent regions remains consistent to prevent bias toward any particular region, while preserving the hierarchical order. With this definition, we can define a density contrast based only on those four values for both long-range  $\delta_{\phi-\rm web}$, and short-range $\delta_{\delta-\rm web}$ components. {\color{black} The cross-correlation between these cosmic-web based fields and the density field are shown in figure \ref{fig:CROSSPK}.} \textcolor{black}{The effect of the $\lambda_{\rm th}$ parameter on these density fields is to make tighter structures with increasing values of $\lambda_{\rm th}$ but always close to zero. Above 0.2 the filamentary structure starts to disappear and the voids occupy most of the volume.}

We propose selecting the threshold value $\lambda_{\rm th}$ in terms of the cross-correlation between the cosmic web density fields $\delta_{\phi-\rm web}$, $\delta_{\delta-\rm web}$, \textcolor{black}{defined from the ALPT density field}, and the \textcolor{black}{UNIT} halo reference catalogue (section \ref{sec:calibration}). After exploring a wide range of threshold values, we have determined that $\lambda_{\rm th} = 0.05$ is the most suitable choice. In figure \ref{fig:CROSSPK}, we display the correlation for $\lambda_{\rm th} = 0.05$, along with lower ($\lambda_{\rm th} = 0.0$) and upper ($\lambda_{\rm th} = 0.1$) limits.

The prominent feature of this plot is that, on small scales, the $\delta$-web exhibits a higher correlation with the reference halos compared to the $\phi$-web. This observation supports the notion that $\delta$-web classification encompasses information on small scales, effectively modeling short-range non-local behavior. Across different thresholds, the cross-correlation with the $\phi$-web remains relatively consistent at both large and small scales, with a slight increase in correlation at intermediate scales for higher $\lambda_{\rm th}$ values (up to 0.1, beyond which it starts to decrease). However, the correlation with the $\delta$-web displays significant differences for various thresholds: at large and intermediate scales, the correlation for $\lambda_{\rm th} = 0.0$ is notably lower compared to other cases. Across almost all scales, the $\delta$-web with $\lambda_{\rm th} = 0.05$ demonstrates the highest correlation with the reference catalogue.
 In light of these findings, we have opted to utilise $\lambda_{\rm th} = 0.05$ in this study.

\begin{figure}[H]
\hspace{-0.6cm}
\begin{tikzpicture}[node distance=4cm]
\node (0) [box0] {Full Halo Sample};

\node [below of=0, node distance=8.4cm] (tot) {};
\node [below of=tot, node distance=1.45cm] (totMU) {\begin{tabular}{c}$\underbrace{\hspace{3cm}_{}}$ \\ \\ $\delta_{\rm tr}$
\end{tabular}};
\node [below of=tot, node distance=1.7cm] (totM) {};
\node [right of=totM, node distance=4.5cm] (phiM) {\begin{tabular}{c}$\underbrace{\hspace{4.cm}_{}}$\\ \\ $\phi$-web\\
  $B_{\rm NL}^{\rm LR}(\delta)$
\end{tabular}};
\node [right of=phiM, node distance=5.05cm] (deltaM) {\begin{tabular}{c}$\underbrace{\hspace{4cm}_{}}$\\ \\
$\delta$-web\\
     $B_{\rm NL}^{\rm SR}(\delta)$
\end{tabular}};
\node [left of=tot, node distance=1.5cm] (totL) {};
\node [right of=tot, node distance=1.5cm] (totR) {};
\node [right of=totR, node distance=1.cm] (phiL) {};
\node [right of=phiL, node distance=4.cm] (phiR) {};
\node [right of=phiR, node distance=1.cm] (deltaL) {};
\node [right of=deltaL, node distance=4.cm] (deltaR) {};

\node (1) [box1, right of=0, yshift=6.5cm, xshift=0.5cm] {\small Knots $\left\lbrace\begin{array}{c} V_{\rm fr}= 2.6 \\ N_{\rm fr}=15.8 \end{array}\right.$ };
\node (2) [box2, right of=0, yshift=2cm, xshift=0.5cm] {\small Filaments $\left\lbrace\begin{array}{c} V_{\rm fr}=25.1 \\ N_{\rm fr}=53.6 \end{array}\right.$};
\node (3) [box3, right of=0, yshift=-2.5cm, xshift=0.5cm] {\small Sheets $\left\lbrace\begin{array}{c} V_{\rm fr}=51.1 \\ N_{\rm fr}=28.2 \end{array}\right.$};
\node (4) [box4, right of=0, yshift=-7.cm, xshift=0.5cm] {\small Voids $\left\lbrace\begin{array}{c} V_{\rm fr}=21.2 \\ N_{\rm fr}=2.4 \end{array}\right.$};

\node (11) [box1, right of=1, yshift=1.5cm, xshift=1cm] {\scriptsize Knots $\left\lbrace\begin{array}{c} V_{\rm fr}=0.7 \\ N_{\rm fr}=6.3 \end{array}\right.$ $n_{\rm p}=2$};

\node [right of=11, node distance=2.cm] (localU) {};
\node [above of=localU, node distance=.5cm] (localUU) {};
\node [below of=localU, node distance=8.2cm] (localM1) {};
\node [right of=localM1, node distance=2.cm] (localM) {$\hspace{-1cm}\begin{rcases}
    \vspace{17cm}
\end{rcases}$
\begin{tabular}{c}
$B_{\rm L}(\delta)$\\$+$\\$B^{\epsilon}(\delta)$
\end{tabular}};

\node (12) [box2, right of=1, yshift=.5cm, xshift=1cm] {\scriptsize Filaments $\left\lbrace\hspace{-0.1cm}\begin{array}{c} V_{\rm fr}=1.2 \\ N_{\rm fr}=7.5 \end{array}\right.$\hspace{-0.1cm} $n_{\rm p}=2$};
\node (13) [box3, right of=1, yshift=-0.5cm, xshift=1cm] {\scriptsize Sheets $\left\lbrace\begin{array}{c} V_{\rm fr}=0.6 \\ N_{\rm fr}=1.9 \end{array}\right.$ $n_{\rm p}=2$};
\node (14) [box4, right of=1, yshift=-1.5cm, xshift=1cm] {\scriptsize Voids $\left\lbrace\begin{array}{c} V_{\rm fr}=0.1 \\ N_{\rm fr}=0.1 \end{array}\right.$ $n_{\rm p}=2$};

\node (21) [box1, right of=2, yshift=1.5cm, xshift=1cm] {\scriptsize Knots $\left\lbrace\begin{array}{c} V_{\rm fr}=0.1 \\ N_{\rm fr}=1.2 \end{array}\right.$ $n_{\rm p}=2$};
\node (22) [box2, right of=2, yshift=.5cm, xshift=1cm] {\scriptsize Filaments \hspace{-0.1cm}$\left\lbrace\hspace{-0.1cm}\begin{array}{c} V_{\rm fr}=4.1 \\ N_{\rm fr}=19.6 \end{array}\right.$\hspace{-0.1cm} $n_{\rm p}=2$};
\node (23) [box3, right of=2, yshift=-0.5cm, xshift=1cm] {\scriptsize  Sheets $\left\lbrace\begin{array}{c} V_{\rm fr}=13.1 \\ N_{\rm fr}=26.1 \end{array}\right.$ $n_{\rm p}=4$};
\node (24) [box4, right of=2, yshift=-1.5cm, xshift=1cm] {\scriptsize  Voids $\left\lbrace\begin{array}{c} V_{\rm fr}=7.8 \\ N_{\rm fr}=6.7 \end{array}\right.$ $n_{\rm p}=4$};

\node (31) [box1, right of=3, yshift=1.5cm, xshift=1cm] {\scriptsize Knots $\left\lbrace\begin{array}{c} V_{\rm fr}<0.1 \\ N_{\rm fr}<0.1 \end{array}\right.$ $n_{\rm p}=0$};
\node (32) [box2, right of=3, yshift=.5cm, xshift=1cm] {\scriptsize Filaments $\left\lbrace\hspace{-0.1cm}\begin{array}{c} V_{\rm fr}=0.1 \\ N_{\rm fr}=0.3 \end{array}\right.$\hspace{-0.1cm} $n_{\rm p}=2$};
\node (33) [box3, right of=3, yshift=-0.5cm, xshift=1cm] {\scriptsize  Sheets $\left\lbrace\begin{array}{c} V_{\rm fr}=8.3 \\ N_{\rm fr}=10.5 \end{array}\right.$ $n_{\rm p}=4$};
\node (34) [box4, right of=3, yshift=-1.5cm, xshift=1cm] {\scriptsize  Voids $\left\lbrace\begin{array}{c} V_{\rm fr}=42.7 \\ N_{\rm fr}=17.4 \end{array}\right.$ $n_{\rm p}=4$};

\node (35) [box1, right of=4, yshift=1.5cm, xshift=1cm] {\scriptsize Knots $\left\lbrace\begin{array}{c} V_{\rm fr}<0.1 \\ N_{\rm fr}<0.1 \end{array}\right.$ $n_{\rm p}=0$};
\node (36) [box2, right of=4, yshift=.5cm, xshift=1cm] {\scriptsize Filaments \hspace{-0.1cm}$\left\lbrace\begin{array}{c} V_{\rm fr}<0.1 \\ N_{\rm fr}<0.1 \end{array}\right.$\hspace{-0.1cm} $n_{\rm p}=0$};
\node (37) [box3, right of=4, yshift=-0.5cm, xshift=1cm] {\scriptsize  Sheets $\left\lbrace\begin{array}{c} V_{\rm fr}<0.1\\ N_{\rm fr}<0.1 \end{array}\right.$ $n_{\rm p}=0$};
\node (38) [box4, right of=4, yshift=-1.5cm, xshift=1cm] {\scriptsize  Voids $\left\lbrace\begin{array}{c} V_{\rm fr}=21.1 \\ N_{\rm fr}=2.3\end{array}\right.$ $n_{\rm p}=2$};

\draw [arrow] (0.east)--++(.5,0) |- (1);
\draw [arrow] (0.east)--++(.5,0) |- (2);
\draw [arrow] (0.east)--++(.5,0) |- (3);
\draw [arrow] (0.east)--++(.5,0) |- (4);
\draw [arrow] (1.east)--++(.5,0) |- (11);
\draw [arrow] (1.east)--++(.5,0) |- (12);
\draw [arrow] (1.east)--++(.5,0) |- (13);
\draw [arrow] (1.east)--++(.5,0) |- (14);
\draw [arrow] (2.east)--++(.5,0) |- (21);
\draw [arrow] (2.east)--++(.5,0) |- (22);
\draw [arrow] (2.east)--++(.5,0) |- (23);
\draw [arrow] (2.east)--++(.5,0) |- (24);
\draw [arrow] (3.east)--++(.5,0) |- (31);
\draw [arrow] (3.east)--++(.5,0) |- (32);
\draw [arrow] (3.east)--++(.5,0) |- (33);
\draw [arrow] (3.east)--++(.5,0) |- (34);
\draw [arrow] (4.east)--++(.5,0) |- (35);
\draw [arrow] (4.east)--++(.5,0) |- (36);
\draw [arrow] (4.east)--++(.5,0) |- (37);
\draw [arrow] (4.east)--++(.5,0) |- (38);
\end{tikzpicture}
 \caption{Hierarchical cosmic web classification of the UNIT halo reference catalogue. The colour-code has been chosen to match the one of the cosmic web classification shown in figure ~\ref{fig:COSMICWEB}. The volume fraction $V_{\rm fr}$ is the percentage number of cells in each region w.r.t to the total box. The number fraction $N_{\rm fr}$ is the percentage number of halos in each region w.r.t to the full sample. The number of bias parameters $n_{\rm p}$ used in each region is also indicated. Since there are only a few halos in voids, two parameters are enough to describe all the $\delta$-web region. Regions with extremely low number densities ($n_{\rm tr}|_{\rm cw}$) can be modelled with a constant bias solely determined by $n_{\rm tr}|_{\rm cw}$. In the bottom of the plot we include the label of both classifications and their relation with the non-local bias terms.}
    \label{fig:tikz}
\end{figure}
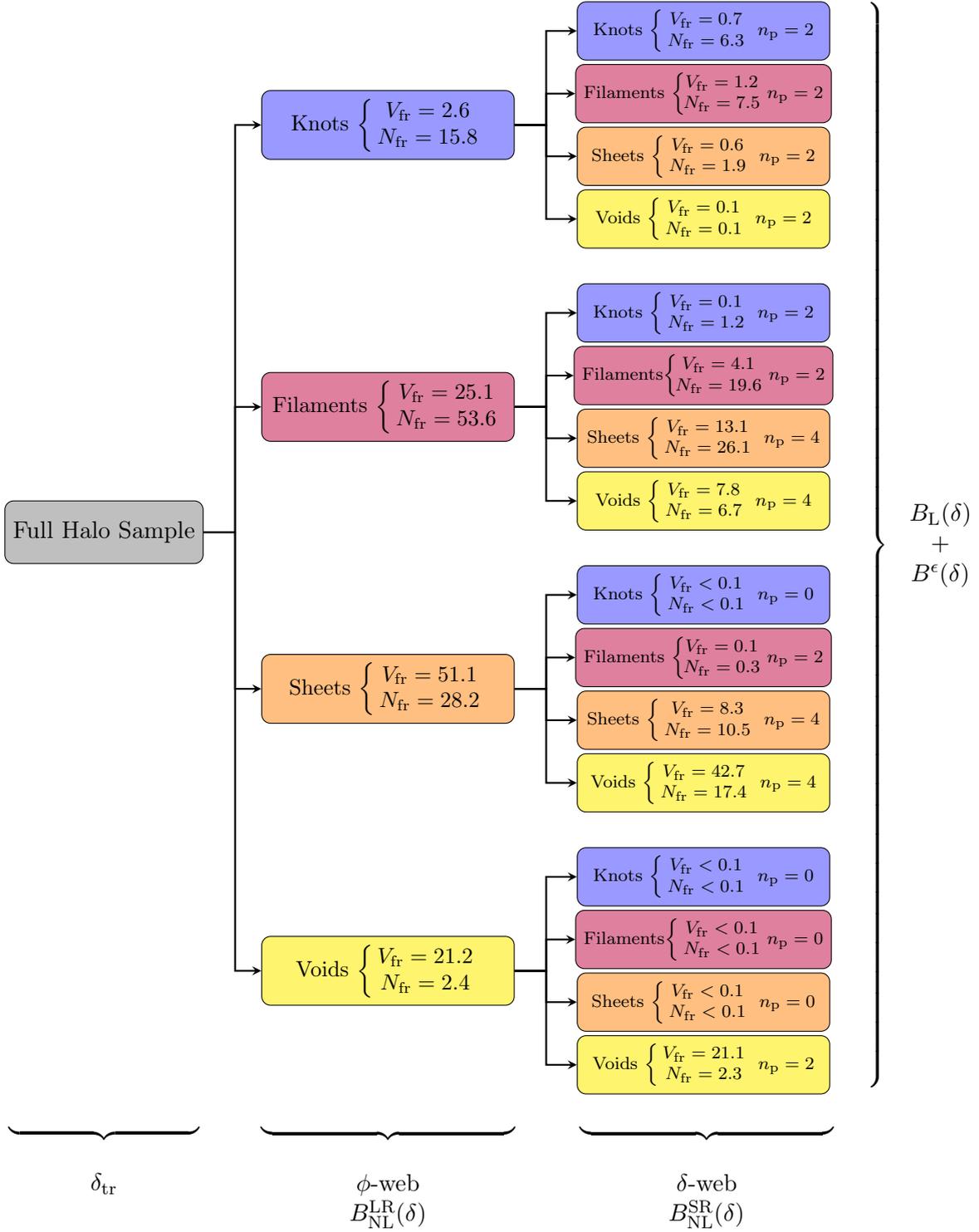

 Moreover, figure \ref{fig:tikz} depicts the hierarchical cosmic web classification scheme for the halo reference catalogue used for calibration in section \ref{sec:calibration}. In the subsequent section, we demonstrate how progressively incorporating $\phi$-web and $\delta$-web information significantly enhances the 1-, 2-, and 3-point statistics, culminating in the reproduction of the reference catalogue with remarkable accuracy.

As a consequence of our hierarchical cosmic web classification, various quantities related to the local deterministic and stochastic bias components have to be computed for each sub-region.  In particular, the bias parameters depend on the cosmic web region denoted by the subscript $\text{cw}$: $\{\alpha|_{\rm cw},{\rho_{\epsilon}}|_{\rm cw}, \epsilon|_{\rm cw},{\rho_{\epsilon}'}|_{\rm cw}, \epsilon'|_{\rm cw}, \beta|_{\rm cw}\}$ and, thus, also the normalisation: $f_{\rm tr}|_{\rm cw}$ depending on the specific bias, volume: $\langle 1+B(\delta_{\rm tr}|\delta)|_{\rm cw}\rangle_{V}$, and mean number density: $n_{\rm tr}|_{\rm cw}$.  {\color{black} We dub the resulting bias model \texttt{HICOBIAN}: {\bf HI}erarchical {\bf CO}smic-web {\bf BIA}s {\bf N}onlocal model.}

\section{Calibration on N-body halo catalog}

\label{sec:calibration}

In this section, we present  the application of the method  to generate halo mock catalogues. 

\subsection{Reference catalog: UNIT simulation}

We employ a gravity-only simulation from the UNIT project\footnote{\url{http://www.unitsims.org}}, specifically a fixed-amplitude run denoted as UNITSIM1 (U1) \citep[][]{Chuang_2019MNRAS}, which is consistent with the cosmological parameters of Planck Collaboration 2016 \citep{Planck_2016A}, i.e., $\Omega_m=0.3089$, $h\equiv \mathrm{H}_0/100=0.6774$, $n_s=0.9667$ and $\sigma_8 = 0.8147$ (see table 4 in \citep{Planck_2016A}). The simulation tracks the dynamic evolution of $4096^3$ dark matter (DM) particles within a periodic box of 1 Gpc\,$h^{-1}$ on a side, starting at scale factor $a \equiv 1/(1 + z) = 0.01$ ($z = 99$) with initial conditions generated with FastPM \citep{Feng_2016MNRAS}. The DM particles undergo evolution up to redshift $z=0$ with the particle-mesh and tree algorithms implemented in the \texttt{GADGET} code \citep{Springel2005MNRAS}. This setup ensures a mass resolution of $1.2 \times 10^9~M_{\odot} h^{-1}$.
Halos were identified with the phase-space halo finder \texttt{ROCKSTAR}\footnote{\url{ https://bitbucket.org/gfcstanford/rockstar}} \citep{Behroozi2013ApJ}. The \texttt{CONSISTENT}\footnote{\url{ https://bitbucket.org/pbehroozi/consistent-trees}} algorithm was applied on top of the \texttt{ROCKSTAR} catalogues  to enhance the consistency of merger histories through the implementation of dynamic tracking for halo progenitors (for details, please refer to \citep{BehrooziConsist2013ApJ}).
As for the reference catalogue used in this research, we focus in the halo distribution at $z=1.032$ with spherical overdensity masses of $M_{200c} \gtrsim 10^{11}~M_\odot\,h^{-1}$. This criterion agrees with the expected mass cut at which dark matter halos can host Emission Line Galaxies (ELGs) (as detailed in studies such as \citep{Alam_2020MNRAS}). 
 For this reason, we have chosen a halo mass cut of $ 10^{11}~M_\odot\,h^{-1}$ yielding a reference catalogue  containing a total of 8,278,311 distinct halos.

\subsection{Mock catalog}

 The first step involves running ALPT with the \texttt{WebON} code \cite{KitauraSinigaglia_2024b} in a cubical volume of 1 $h^{-1}$Gpc side to $z=1.032$. To this end, we employ a down-sampled version  of the original initial conditions (corresponding to the UNITSIM1 simulation relying on $4096^3$ cells and particles) to a mesh of 256$^3$ cells and particles. This results in a cell side resolution of approximately 3.9 $h^{-1}$ Mpc. We consistently use the same cosmology as the UNITSIM1. We define the dark matter density contrast after applying the CIC mass assignment scheme combined with tetrahedra-tesselation \citep{Abel_2012}.
 We then use the resulting dark matter field to calibrate the bias parameters.   Subsequently, we execute 100 realisations with random initial conditions to ensure robust statistical analysis. 

Furthermore, from the \texttt{WebON} code, we obtain the $\phi$-web and $\delta$-web classifications for the ALPT dark matter density field using a threshold value of $\lambda_{\rm th}=0.05$. This classification is also applied to the halo reference catalogue to determine the number density in each cosmic web region. The classification scheme is illustrated in figure \ref{fig:tikz}, where the two numbers $V_{\rm fr}$ and $N_{\rm fr}$ represent the volume and number fraction, respectively, characterising each region of the cosmic web.

In the second step, we apply the bias model described in \eqref{eq:deterministic_bias} to each of the individual regions. Although initially there are six possible free parameters $\{\alpha, \beta, \rho_{\epsilon}, \epsilon, \rho^{\prime}_{\epsilon}, \epsilon^{\prime}\}$, we find that we can successfully fit most of the individual regions using only two parameters, $\{\alpha, \beta\}$. For some regions, we require four parameters by incorporating the exponential damping for high-density regions, but in neither case do we need all six parameters. \textcolor{black}{The number of parameters used in each region is labeled as $n_{\rm p}$ in figure \ref{fig:tikz}. In this work, in contrast to \citep{Kitaura_2014}, we do not need the low-density threshold bias term of \eqref{eq:deterministic_bias}. This is accounted for by classifying the short-range cosmic web, which leads to varying bias parameters and number densities in each region. This classification controls the emergence of haloes in expanding low-density regions.}

We are conducting separate studies to investigate which parameters are relevant for each galaxy type.

\subsection{Results}

\begin{figure}
    \centering
    \includegraphics[scale=0.19]{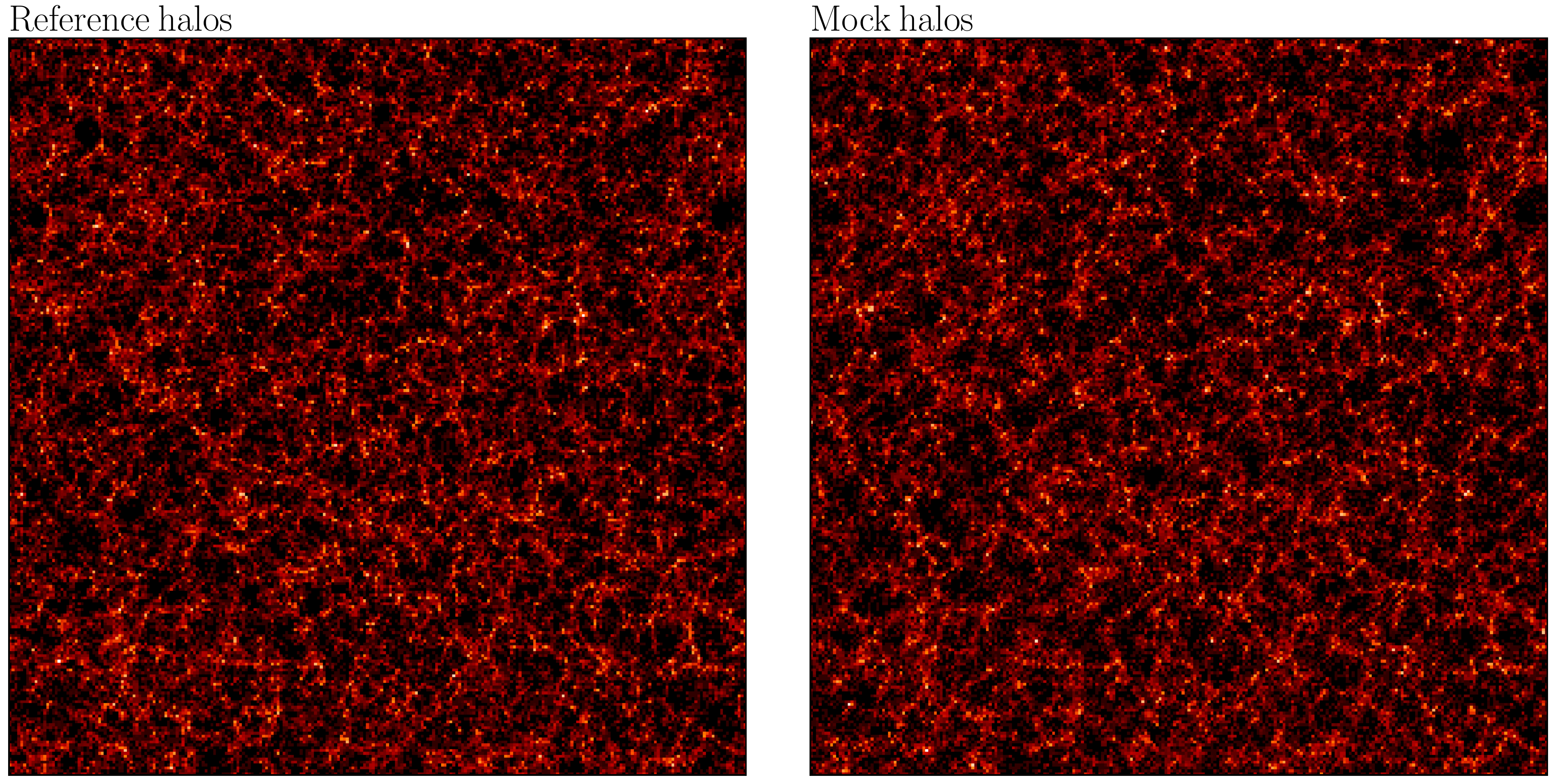}
    \caption{Visual comparison of the reference and a mock halo catalog. The slices are $\sim$20 $h^{-1}$ Mpc thick and 1000 $h^{-1}$ Mpc side. The logarithm of the halo density contrast $\delta_{\rm tr}$ is shown and darker regions are related to lower densities. Both fields share the same initial conditions down-sampled to  resolutions of $\sim3.9\,h^{-1}$ Mpc cell side. Below that resolution, stochastic bias dominates and the halo distributions differ.}
    \label{fig:visualcomparison}
\end{figure}

Figure \ref{fig:visualcomparison} depicts a visual comparison between the reference and the mock halo catalogue. The disparities observed stem from stochasticity induced by the various realisations below the scale of the down-sampled field. Nevertheless, both fields exhibit striking similarities, and the large-scale structure is clearly discernible and consistent in both panels.

\begin{figure}[H]
    \centering
    \begin{tabular}{c}
    \hspace{-.4cm}
    \includegraphics[scale=0.19]{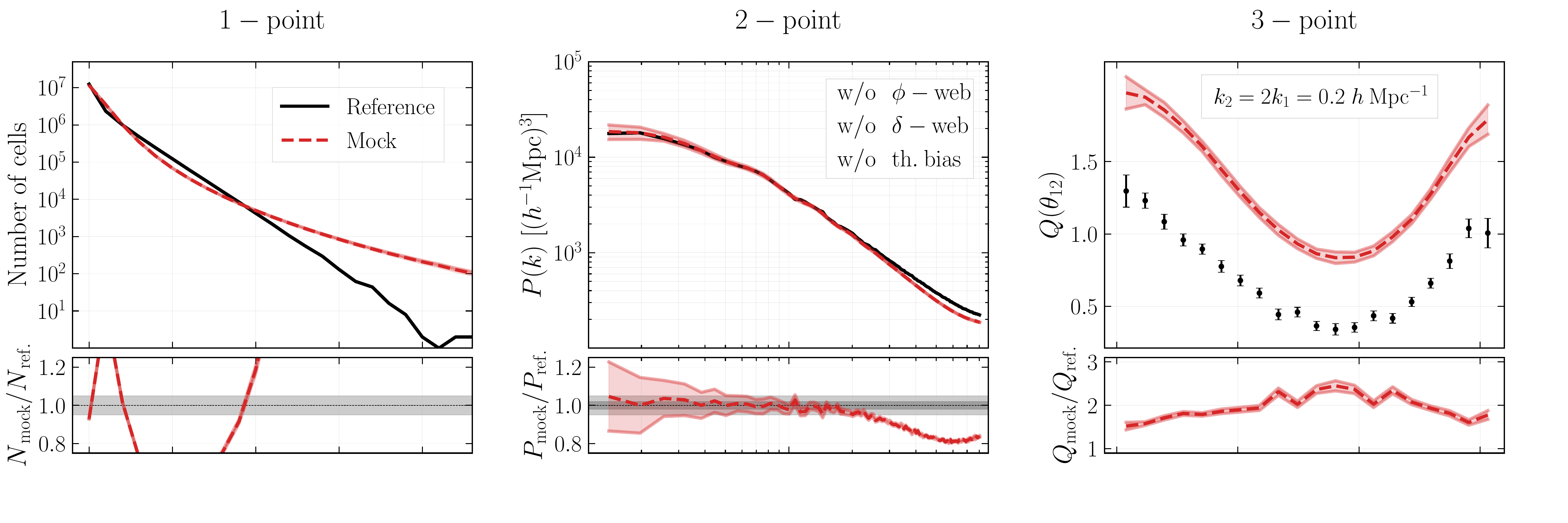}\vspace{-.7cm}\\
    \hspace{-.4cm}
     \includegraphics[scale=0.19]{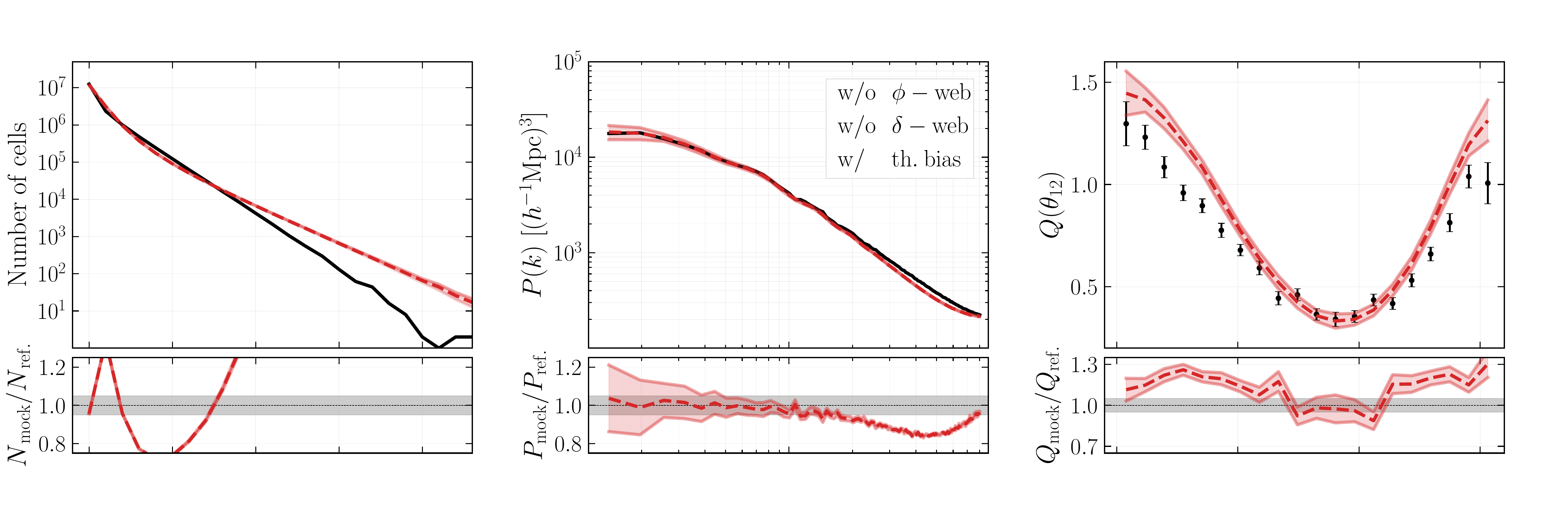}\vspace{-.7cm}\\
     \hspace{-.4cm}
    \includegraphics[scale=0.19]{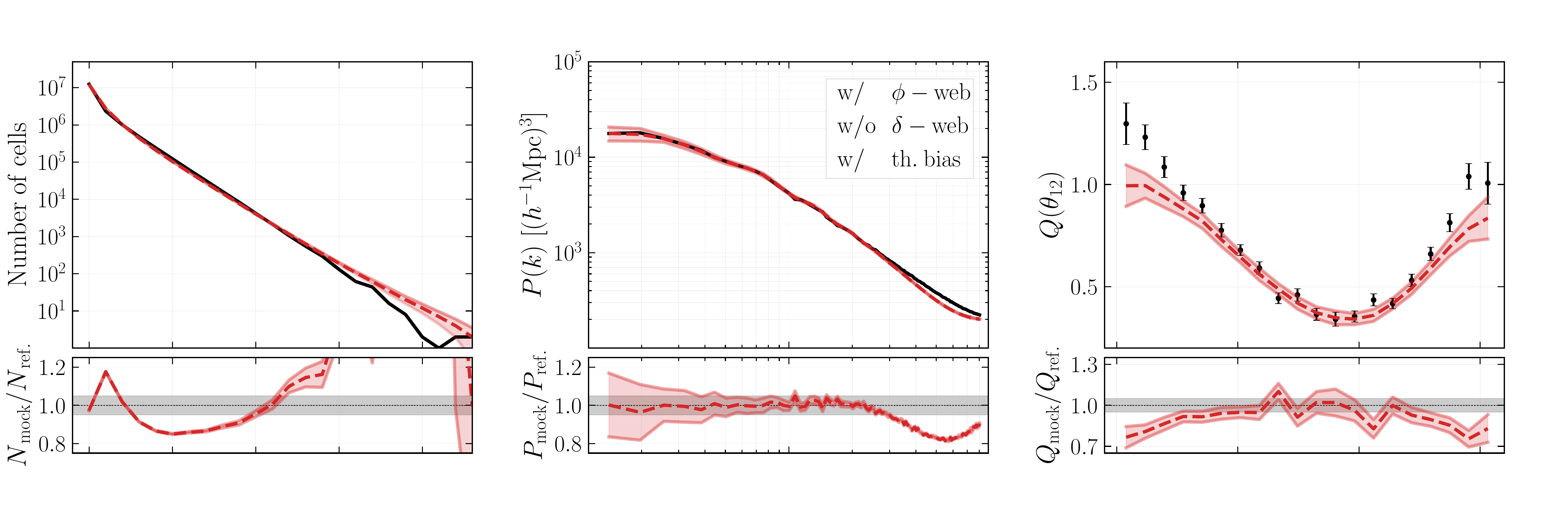}\vspace{-.7cm}\\
    \hspace{-.4cm}
    \includegraphics[scale=0.19]{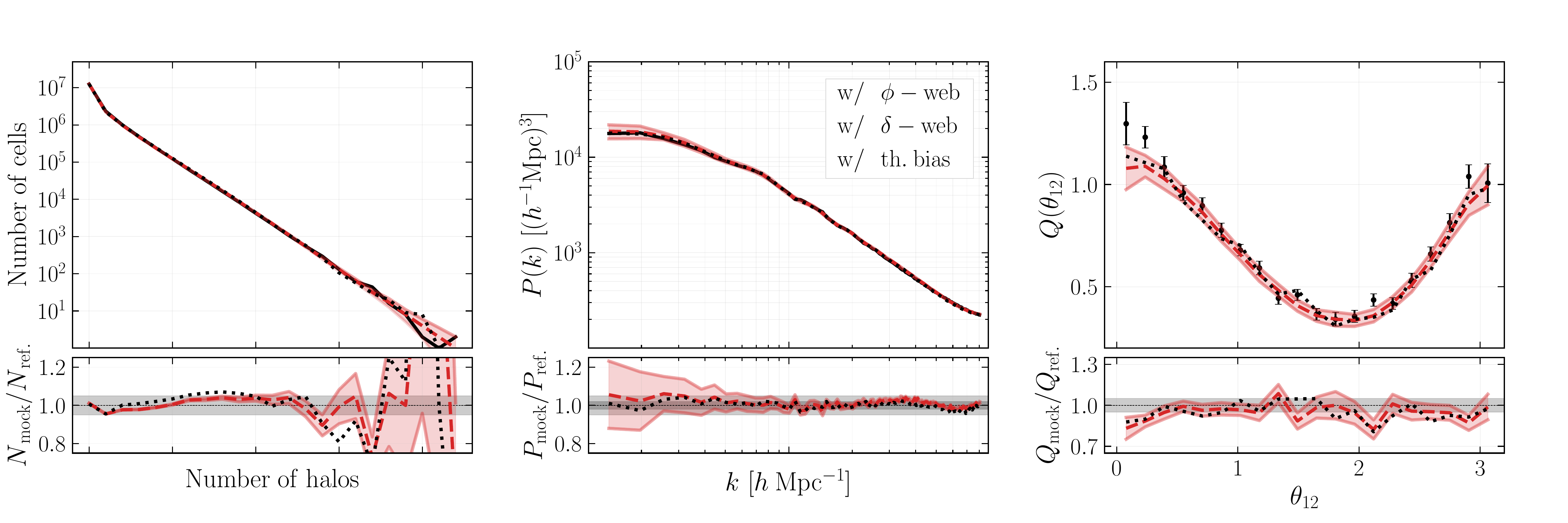}
    \end{tabular}
    \caption{Statistical comparison of the UNIT simulation reference catalogue (black) and the different mock catalogues (red). The probability density function PDF (left column), the power-spectrum $P(k)$ (middle) and the reduced bispectrum $Q(\theta_{12})$ (right) are shown for a model describing a deterministic and stochastic bias model (first row), to which we add a threshold bias (second) and then, progressively, the $\phi$-web and $\delta$-web information (third and fourth rows). The dashed red line is the mean over 100 realisations and the shaded area represents the standard deviation. The grey area in the subsets represent the 5 percent deviation (2 percent also included in the bottom row). The error bars on the reference bispectrum come from the 100 realizations.}
    \label{fig:STATISTICS}
\end{figure}

A quantitative analysis is presented in figure \ref{fig:STATISTICS}, where each row illustrates the 1-, 2- and 3-point statistics or, equivalently, the probability density function PDF (left), the power-spectrum $P(k)$ (middle), and the reduced bispectrum $Q(\theta_{12})$ (right) for the various bias models considered in this study. We progressively include more bias terms from the top to the bottom rows.
The summary statistics corresponding to the reference catalogue is represented with the black solid lines or black dots and corresponding error bars. \textcolor{black}{The mock catalogue mean over 100 realisations is represented with the red dashed line and the red shaded area corresponds to the standard deviation region.}

The PDF \citep{Uhlemann_2020}, the power-spectrum \citep{Alam_2017} and the bispectrum \citep{Sefusatti_2006, Sugiyama_2023} are powerful tools for extracting cosmological information from the large-scale structure of the Universe by constraining cosmological parameters. The PDF is understood as the number of cells containing a given number of objects. The Fourier-space 2- and 3-point functions, i.e., the power-spectrum and bispectrum,  respectively, are defined as
\begin{eqnarray}
    \label{eq:bispectrum}
\langle \delta_{\rm tr}(\boldsymbol{k}) \delta_{\rm tr}(\boldsymbol{k}') \rangle &\equiv& (2\pi)^3 P(k) \delta^D(\boldsymbol{k} - \boldsymbol{k}')\,,\\
\langle \delta_{\rm tr}(\boldsymbol{k}_1) \delta_{\rm tr}(\boldsymbol{k}_2) \delta_{\rm tr}(\boldsymbol{k}_3) \rangle &\equiv& (2\pi)^3 B(k_1, k_2, k_3) \delta^D(\boldsymbol{k}_1 + \boldsymbol{k}_2 + \boldsymbol{k}_3)\,,
\end{eqnarray}
where the ensemble average stands for $k$-shell averages, while the reduced bispectrum is defined as
\begin{equation}
Q(\theta_{12}|k_1, k_2) \equiv \frac{B(k_1, k_2,k_3)}{P(k_1)P(k_2) + P(k_2)P(k_3) + P(k_1)P(k_3)}\,.
\end{equation}
We have focused on a particular configuration of the reduced bispectrum, $k_2=2k_1=0.2\, h$ Mpc$^{-1}$, which is typically studied, as it is relevant to BAO and RSD analyses. We are using larger volume simulations based on the AbacusSummit \citep{Maksimova_2021} simulation suite  to thoroughly study a larger variety of configurations for halos and each galaxy type tracer relevant to the DESI and EUCLID collaborations. Such studies will be presented in forthcoming papers. 
The bispectrum projection is shown as a function of the angle $\theta_{12}$ between the wavevectors $\vb*{k}_1$ and $\vb*{k}_2$, given by $k_3^{2} = k_2^2\sin^2\theta_{12} + (k_2\cos\theta_{12} + k_1)^2$.\footnote{We are grateful to Cheng Zhao for making his bispectrum computation code available as presented in \cite{Zhao_2021}.}

The resulting mock catalogues are based on the following bias models (figure \ref{fig:STATISTICS}):
\begin{enumerate}
    \item First row: Local power-law bias 
    \item Second row: Local power-law bias combined with threshold bias 
    \item Third row: Local power-law bias combined with threshold bias and long-range cosmic web classification based on the gravitational potential ($\phi$-web)
    \item Fourth row: Local power-law bias combined with threshold bias and both long-range and short-range cosmic web classification based on the gravitational potential ($\phi$-web) and the density field ($\delta$-web).
\end{enumerate}

As outlined in previous studies \citep{Vakili_2017,Sinigaglia_2024}, the determination of the bias parameters is performed jointly by \textcolor{black}{maximizing} the likelihood of the 1-point PDF and the power-spectrum. This joint optimisation is crucial as the skewness of the PDF is constrained by the bispectrum, aiding in breaking degeneracies \citep{Kitaura_2015}. In particular, we determine the optimal values for our parameters by sampling their posterior distribution through the affine-invariant \texttt{emcee} Markov Chain Monte Carlo (MCMC) sampler \citep{GoodmanWeare_2010,ForemanMackey_2013}. 
{\color{black} The total log-likelihood is given by $-2\ln(\mathcal{L})=-2\left[\ln(\mathcal{L})_{P(k)}({\rm ref}|\theta)+\ln(\mathcal{L})_{\rm PDF}({\rm ref}|\Theta)\right]$, where "ref" denotes the reference, and $\ln(\mathcal{L})_{P(k)}({\rm ref}|\theta)$ and $\ln(\mathcal{L})_{\rm ref}({\rm ref}|\theta)$ are Gaussian likelihoods (see also Eqs. 6--9 in \cite{Vakili_2017}). We assume no correlation between the PDF and the $P(k)$. Also, we assume diagonal covariance matrices, with zero off-diagonal components and diagonal elements given by $\sigma_{P(k)}^2=4\pi P_{\rm ref}(k)/V_{\rm box} k \Delta k $ and $\sigma_{\rm PDF}^2=N_n$ -- with $N_n$ the number of cells containing $n$ halos -- for the power-spectrum and the PDF, i.e. Gaussian and Poisson uncertainties, respectively.}

 { The power-spectrum likelihood assumes a conservative overestimated variance, since the reference and the calibration catalogues share the same fixed-amplitude and down-sampled initial conditions. However, one has to consider that the down-sampling process introduces stochastic bias, which has some impact on the power-spectrum at low modes. In fact, this variance is lower than additionally allowing for random initial conditions. Nonetheless, we have checked that this likelihood does not introduce large-scale biases (see calibration dotted lines in the lowest panel of figure \ref{fig:STATISTICS}). The calibration shows a better agreement than the mean of one hundred random realisations in the power-spectrum.}

\textcolor{black}{The PDF is fitted in the range [0,20], for the number of halos, and the power-spectrum in the wavenumber range [0,0.8] $h$ Mpc$^{-1}$. The free parameters are $\theta=\{\alpha, \beta\}$ or $\theta=\{\alpha, \beta, \rho_{\epsilon}, \epsilon\}$ depending on the region.}

Through our approach, it is evident how we progressively enhance the fit to the 1-, 2-, and 3-point statistics.
 With the simplest model, we are significantly divergent from reproducing the reference catalogue. There are substantial deviations in the PDF and the bispectrum, and the power-spectrum fails particularly towards high wave numbers $k$ (small scales). 

The second row of plots demonstrates how incorporating a threshold bias \eqref{eq:deterministic_bias} improves statistics, notably by reducing the significant deviation in the bispectrum. However, the statistics are still not well reproduced. Therefore, we decided to incorporate information on the different cosmic web environments (knots, filaments, voids, and sheets) by individually applying the bias model to each region rather than directly to the entire catalogue. This leads to significant improvements (third row): the PDF is much closer to the reference, the power-spectrum is reproduced within a 5 percent deviation until $k=0.3 \, h$ Mpc$^{-1}$, and only the wings of the bispectrum are not entirely compatible with the UNIT simulation catalogue.

Despite these improvements, we aim for percentage accuracy and ultimately achieve it by incorporating small-scale information through hierarchical cosmic web classification. Now, the four regions of the $\phi$-web are subclassified based on the density field tensor, which, as demonstrated in section \ref{sec:non-local bias}, indirectly models the short-range non-local bias. The bias model \eqref{eq:deterministic_bias} is now applied to each of the 16 regions, and the statistical result of combining them to create the entire catalogue is shown in the fourth row of figure \ref{fig:STATISTICS}. The PDF is now reproduced with 5 percent accuracy, he power-spectrum is compatible within 2 percent deviations almost up to $k=0.8 \, h$ Mpc$^{-1}$, and the bispectrum, including the wings, is reproduced within error bars.

We find that the low-density threshold bias, previously required in studies utilising the \texttt{PATCHY} code, becomes unnecessary. This is because the short-range $\delta$-web classification already provides a comprehensive enough description of halo formation physics at the scales analysed in our work. As a consequence, the hierarchical cosmic web classification allows us to dramatically reduce the number of parameters $n_{\rm p}$ used in each region, so that the total number of parameters employed in our study is of 32 instead of  96 ($=16\text{ (regions) }\times 6 \text{ (parameters)}$), as indicated in figure \ref{fig:STATISTICS}. 

\textcolor{black}{We study the goodness of the fits in terms of the $\chi^2$ test, as shown in table \ref{tab:goodness},  quantifying the fits shown in figure \ref{fig:STATISTICS}. Note that the bispectrum is not used in the fit but for a completeness of the analysis.} A decrease in the $\chi^2$, or equivalently an improvement of the fit accuracy, is appreciated when including more bias terms progressively from the first to the fourth row. Important deviations in the PDF, the power-spectrum and the bispectrum are present in the fit until the non-local bias is included through the hierarchical cosmic web leading a final

 remarkable result in the three statistics, $\chi^2 \sim$0, with rigorously accuracy in the two- and three-point functions.

{ Given that we use the same down-sampled initial conditions, we expect to obtain very close results on large scales to the reference catalogues in all summary statistics. Strictly speaking, the reference is from fixed-amplitude initial conditions, so the noise of the reference is difficult to quantify. Hence, the chi-squared values are not faithfully quantifying how well the mock matches with the reference, instead, they serve as quantitative references about how much the clustering statistics are improved with different statistics of the density field. }

It should be emphasised once more that the strength of this approach lies in the ability to treat each subset of the halo catalogue, corresponding to a particular sub-region of the cosmic web, as an independent catalogue. This parallels the practice in clustering analyses where galaxies are divided into categories such as blue and red galaxies. The accuracy of our approach is undoubtedly constrained by the availability of sufficient statistics to effectively constrain the bias parameters of each sub-region. 

{
\begin{table}[ht!]
\centering
\color{black}
\begin{tabular}{ll|cccccccc|}
\cline{3-10}
 &
   &
   
  \multicolumn{1}{l}{} &
  \multicolumn{1}{l|}{First row} &
  \multicolumn{1}{l}{} &
  \multicolumn{1}{l|}{Second row} &
  \multicolumn{1}{l}{} &
  \multicolumn{1}{l|}{Third row} &
  \multicolumn{1}{l}{} &
  \multicolumn{1}{l|}{Fourth row} \\ \hline
\multicolumn{1}{|c|}{}         & PDF         &  & 12564.10 &  & 1346.96 &  & 29.76 &  & 1.10 \\ \cline{2-2}
\multicolumn{1}{|l|}{$\chi^2$} & $P(k)$      &  & 2.32     &  & 1.47    &  & 1.39  &  & 0.05 \\ \cline{2-2}
\multicolumn{1}{|l|}{}         & $Q(\theta_{12})$ &  & 19.99    &  & 0.57    &  & 0.31  &  & 0.13 \\ \hline
\end{tabular}
\caption{\color{black}Goodness of fit in terms of the $\chi^2$ parameter for the PDF, the power-spectrum and the reduced bispectrum in each of the rows shown in figure \ref{fig:STATISTICS}. }
\label{tab:goodness}
\end{table}
}

Nonetheless, this approach is less susceptible to over-fitting compared to machine learning techniques, as the model (and its degrees of freedom) is constrained by its analytical formulation for both the deterministic and stochastic bias components. One crucial advantage over the \texttt{BAM} approach is that it does not require a kernel to reproduce the power-spectrum. This capability enables us to efficiently apply our field-level bias mapping technique in redshift shells to a dark matter lightcone, such as the ones we have computed with the \texttt{WebON} code.

We are exploring in separate studies transformations of the density field which improve the description of the cosmic web towards smaller scales and allows us to explore such scales in the 2- and 3-point statistics.

\section{Summary and discussion}
\label{summary}
In this study, we have introduced an innovative approach for generating mock catalogues which extends the deterministic and stochastic bias model by incorporating a hierarchical classification of the cosmic web. This augmentation allows us to effectively model non-local bias, ensuring the positive definiteness of the expected number counts of tracers.

By integrating information from the cosmic web classification, specifically the $\phi$-web and $\delta$-web, we significantly enhance the fidelity of our bias model. This comprehensive framework enables us to capture the influence of large-scale structure on tracer distributions and accurately reproduce statistical properties such as the probability density function, power-spectrum and bispectrum.

One of the most remarkable findings of this study is the simplification of the bias model as we incorporate the hierarchical cosmic web classification. We observe that each region can typically be modelled with just two parameters, and in some instances, only four parameters are required. This reduction in the number of parameters not only streamlines the modelling process but also enhances the interpretability and efficiency of our approach. It underscores the effectiveness of leveraging cosmic web information to capture the complexity of large-scale structure with minimal complexity in the bias model.

Our method offers several key advantages, including its computational efficiency, high accuracy in reproducing reference statistics up to third order, and versatility in its application to different tracers of large-scale structure. In comparison to automatic non-parametric learning methods, e.g., \citep{Balaguera_2018}, our approach offers the advantage of being less susceptible to overfitting. This is because our bias model, including the stochastic component, is analytical and determined by only a few parameters.

However, further investigation is needed in future work to verify the robustness of the model assumptions, such as the specific deviation from Poissonity set by the negative binomial distribution. This ongoing research will help to ensure the reliability and validity of our approach across a range of cosmological scenarios and observational data sets.

Also, we conducted a series of 200 large-volume light-cone dark matter simulations using the \texttt{WebON} code within cubical volumes of ($10\,h^{-1}$ Gpc)$^3$ \cite{KitauraSinigaglia_2024b}. We are currently preparing a series of papers presenting mock galaxy catalogues generated using this method with various tracers of the large-scale structure. We also plan to explore the 4-point statistics of our mock catalogues. Previous studies have shown that accurately reproducing the 1-, 2-, and 3-point statistics lays the foundation for reproducing covariance matrices \citep{Baumgarten_2018}. Therefore, we believe that we are on the right track with our current approach to become accurate in the higher-order statistics. The hierarchical cosmic web classification presented in this study could potentially be applied for environmental studies of galaxies to study assembly bias. The assignment methods of halo or galaxy properties can benefit from the findings of this paper extending previous approaches with the short-range classification scheme (see \cite{Zhao_2015}). The combination of this bias model with effective field theory can be as straightforward as considering different local bias parameters at various hierarchical cosmic web regions.

Looking ahead, we anticipate further enhancements by continuing to refine the cosmic web classification with additional terms, such as those based on velocity bias components obtained with fast gravity solvers (e.g., \citep{Kitaura_2024}). Additionally, we plan to implement a nonlinear mapping of the density field to imprint missing power from the approximate low-resolution gravity solver and improve the description towards small scales.

This ongoing development promises to advance our understanding of the complex interplay between cosmic web environments and tracer distributions in the Universe. By incorporating these advancements, we aim to further refine our mock catalogue generation process and utilise this framework for field-level inference. This will potentially lead to even more accurate representations of large-scale structure and facilitate deeper insights into cosmological phenomena.

\section*{Acknowledgements}
We thank the referee for the careful revision of the manuscript. JMCN, FSK, JGF, FS, and GF acknowledge  the Spanish Ministry of Economy and Competitiveness (MINECO) for financing the \texttt{Big Data of the Cosmic Web} project: 
PID2020-120612GB-I00/AEI/10.13039/ 501100011033 under which this work has been conceived and carried out, and the IAC for continuous support to the \texttt{Cosmology with LSS probes} project. JMCN thanks the Research Summer Grant at the IAC for continuing the study of the master thesis and the ICE for support during the writing phase of this paper. JEGF is supported by the Spanish Ministry of Universities, through a Mar\'ia Zambrano grant (program 2021-2023) at Universidad de La Laguna with reference UP2021-022, funded within the European Union-Next Generation EU. FS acknowledges the support of the Swiss National Science Foundation (SNSF) 200021\_214990/1 grant. GF thanks support from the IJC2020-044343-I grant. DFS thanks support from the Swiss National Science Foundation (SNF) 200020\_175751 and 200020\_207379. 

\bibliographystyle{JHEP}
\bibliography{references}

\providecommand{\href}[2]{#2}\begingroup\raggedright\begin{thebibliography}{100}

\bibitem{levi2013desi}
M.~Levi, C.~Bebek, T.~Beers, R.~Blum, R.~Cahn, D.~Eisenstein et~al., \emph{The
  desi experiment, a whitepaper for snowmass 2013},  2013.

\bibitem{laureijs2011euclid}
R.~Laureijs, J.~Amiaux, S.~Arduini, J.L.~Auguères, J.~Brinchmann, R.~Cole
  et~al., \emph{Euclid definition study report},  2011.

\bibitem{Bernardeau_2002}
F.~Bernardeau, S.~Colombi, E.~Gazta{\~{n}}aga and R.~Scoccimarro,
  \emph{Large-scale structure of the universe and cosmological perturbation
  theory}, \href{https://doi.org/10.1016/s0370-1573(02)00135-7}{\emph{Physics
  Reports} {\bfseries 367} (2002) 1}.

\bibitem{Bond_1996}
J.R.~{Bond} and S.T.~{Myers}, \emph{{The Peak-Patch Picture of Cosmic Catalogs.
  I. Algorithms}}, \href{https://doi.org/10.1086/192267}{\emph{\apjs}
  {\bfseries 103} (1996) 1}.

\bibitem{Bond_1996b}
J.R.~{Bond} and S.T.~{Myers}, \emph{{The Peak-Patch Picture of Cosmic Catalogs.
  II. Validation}}, \href{https://doi.org/10.1086/192268}{\emph{\apjs}
  {\bfseries 103} (1996) 41}.

\bibitem{Bond_1996c}
J.R.~{Bond} and S.T.~{Myers}, \emph{{The Peak-Patch Picture of Cosmic Catalogs.
  III. Application to Clusters}},
  \href{https://doi.org/10.1086/192269}{\emph{\apjs} {\bfseries 103} (1996)
  63}.

\bibitem{Stein_2019}
G.~{Stein}, M.A.~{Alvarez} and J.R.~{Bond}, \emph{{The mass-Peak Patch
  algorithm for fast generation of deep all-sky dark matter halo catalogues and
  its N-body validation}},
  \href{https://doi.org/10.1093/mnras/sty3226}{\emph{\mnras} {\bfseries 483}
  (2019) 2236} [\href{https://arxiv.org/abs/1810.07727}{{\ttfamily
  1810.07727}}].

\bibitem{Monaco_2002}
P.~Monaco, T.~Theuns, G.~Taffoni, F.~Governato, T.~Quinn and J.~Stadel,
  \emph{Predicting the number, spatial distribution, and merging history of
  dark matter halos}, \href{https://doi.org/10.1086/324182}{\emph{The
  Astrophysical Journal} {\bfseries 564} (2002) 8}.

\bibitem{Monaco_2013}
P.~{Monaco}, E.~{Sefusatti}, S.~{Borgani}, M.~{Crocce}, P.~{Fosalba},
  R.K.~{Sheth} et~al., \emph{{An accurate tool for the fast generation of dark
  matter halo catalogues}},
  \href{https://doi.org/10.1093/mnras/stt907}{\emph{\mnras} {\bfseries 433}
  (2013) 2389} [\href{https://arxiv.org/abs/1305.1505}{{\ttfamily 1305.1505}}].

\bibitem{Scoccimarro_2002}
R.~{Scoccimarro} and R.K.~{Sheth}, \emph{{PTHALOS: a fast method for generating
  mock galaxy distributions}},
  \href{https://doi.org/10.1046/j.1365-8711.2002.04999.x}{\emph{\mnras}
  {\bfseries 329} (2002) 629}
  [\href{https://arxiv.org/abs/astro-ph/0106120}{{\ttfamily
  astro-ph/0106120}}].

\bibitem{Manera_2013}
M.~{Manera}, R.~{Scoccimarro}, W.J.~{Percival}, L.~{Samushia}, C.K.~{McBride},
  A.J.~{Ross} et~al., \emph{{The clustering of galaxies in the SDSS-III Baryon
  Oscillation Spectroscopic Survey: a large sample of mock galaxy catalogues}},
  \href{https://doi.org/10.1093/mnras/sts084}{\emph{\mnras} {\bfseries 428}
  (2013) 1036} [\href{https://arxiv.org/abs/1203.6609}{{\ttfamily 1203.6609}}].

\bibitem{Manera_2015}
M.~{Manera}, L.~{Samushia}, R.~{Tojeiro}, C.~{Howlett}, A.J.~{Ross},
  W.J.~{Percival} et~al., \emph{{The clustering of galaxies in the SDSS-III
  Baryon Oscillation Spectroscopic Survey: mock galaxy catalogues for the
  low-redshift sample}},
  \href{https://doi.org/10.1093/mnras/stu2465}{\emph{\mnras} {\bfseries 447}
  (2015) 437} [\href{https://arxiv.org/abs/1401.4171}{{\ttfamily 1401.4171}}].

\bibitem{Kitaura_2014}
F.-S.~Kitaura, G.~Yepes and F.~Prada, \emph{{Modelling baryon acoustic
  oscillations with perturbation theory and stochastic halo biasing}},
  \href{https://doi.org/10.1093/mnrasl/slt172}{\emph{Monthly Notices of the
  Royal Astronomical Society: Letters} {\bfseries 439} (2013) L21}
  [\href{https://arxiv.org/abs/https://academic.oup.com/mnrasl/article-pdf/439/1/L21/3421996/slt172.pdf}{{\ttfamily
  https://academic.oup.com/mnrasl/article-pdf/439/1/L21/3421996/slt172.pdf}}].

\bibitem{Kitaura_2015}
F.-S.~{Kitaura}, H.~{Gil-Mar{\'\i}n}, C.G.~{Sc{\'o}ccola}, C.-H.~{Chuang},
  V.~{M{\"u}ller}, G.~{Yepes} et~al., \emph{{Constraining the halo bispectrum
  in real and redshift space from perturbation theory and non-linear stochastic
  bias}}, \href{https://doi.org/10.1093/mnras/stv645}{\emph{\mnras} {\bfseries
  450} (2015) 1836} [\href{https://arxiv.org/abs/1407.1236}{{\ttfamily
  1407.1236}}].

\bibitem{Kitaura_2016}
F.-S.~{Kitaura}, S.~{Rodr{\'\i}guez-Torres}, C.-H.~{Chuang}, C.~{Zhao},
  F.~{Prada}, H.~{Gil-Mar{\'\i}n} et~al., \emph{{The clustering of galaxies in
  the SDSS-III Baryon Oscillation Spectroscopic Survey: mock galaxy catalogues
  for the BOSS Final Data Release}},
  \href{https://doi.org/10.1093/mnras/stv2826}{\emph{\mnras} {\bfseries 456}
  (2016) 4156} [\href{https://arxiv.org/abs/1509.06400}{{\ttfamily
  1509.06400}}].

\bibitem{Vakili_2017}
M.~{Vakili}, F.-S.~{Kitaura}, Y.~{Feng}, G.~{Yepes}, C.~{Zhao}, C.-H.~{Chuang}
  et~al., \emph{{Accurate halo-galaxy mocks from automatic bias estimation and
  particle mesh gravity solvers}},
  \href{https://doi.org/10.1093/mnras/stx2184}{\emph{\mnras} {\bfseries 472}
  (2017) 4144} [\href{https://arxiv.org/abs/1701.03765}{{\ttfamily
  1701.03765}}].

\bibitem{Chuang_2014}
C.-H.~Chuang, F.-S.~Kitaura, F.~Prada, C.~Zhao and G.~Yepes, \emph{Ezmocks:
  extending the zel’dovich approximation to generate mock galaxy catalogues
  with accurate clustering statistics},
  \href{https://doi.org/10.1093/mnras/stu2301}{\emph{Monthly Notices of the
  Royal Astronomical Society} {\bfseries 446} (2014) 2621–2628}.

\bibitem{Zhao_2021}
C.~{Zhao}, C.-H.~{Chuang}, J.~{Bautista}, A.~{de Mattia}, A.~{Raichoor},
  A.J.~{Ross} et~al., \emph{{The completed SDSS-IV extended Baryon Oscillation
  Spectroscopic Survey: 1000 multi-tracer mock catalogues with redshift
  evolution and systematics for galaxies and quasars of the final data
  release}}, \href{https://doi.org/10.1093/mnras/stab510}{\emph{\mnras}
  {\bfseries 503} (2021) 1149}
  [\href{https://arxiv.org/abs/2007.08997}{{\ttfamily 2007.08997}}].

\bibitem{Avila:2014nia}
S.~Avila, S.G.~Murray, A.~Knebe, C.~Power, A.S.G.~Robotham and
  J.~Garcia-Bellido, \emph{{HALOGEN: A tool for fast generation of mock halo
  catalogues}}, \href{https://doi.org/10.1093/mnras/stv711}{\emph{Mon. Not.
  Roy. Astron. Soc.} {\bfseries 450} (2015) 1856}
  [\href{https://arxiv.org/abs/1412.5228}{{\ttfamily 1412.5228}}].

\bibitem{Izard_2016}
A.~{Izard}, M.~{Crocce} and P.~{Fosalba}, \emph{{ICE-COLA: towards fast and
  accurate synthetic galaxy catalogues optimizing a quasi-N-body method}},
  \href{https://doi.org/10.1093/mnras/stw797}{\emph{\mnras} {\bfseries 459}
  (2016) 2327} [\href{https://arxiv.org/abs/1509.04685}{{\ttfamily
  1509.04685}}].

\bibitem{Balaguera_2018}
A.~Balaguera-Antolínez, F.-S.~Kitaura, M.~Pellejero-Ibáñez, C.~Zhao and
  T.~Abel, \emph{{BAM: bias assignment method to generate mock catalogues}},
  \href{https://doi.org/10.1093/mnrasl/sly220}{\emph{Monthly Notices of the
  Royal Astronomical Society: Letters} {\bfseries 483} (2018) L58}
  [\href{https://arxiv.org/abs/https://academic.oup.com/mnrasl/article-pdf/483/1/L58/54700852/mnrasl\_483\_1\_l58.pdf}{{\ttfamily
  https://academic.oup.com/mnrasl/article-pdf/483/1/L58/54700852/mnrasl\_483\_1\_l58.pdf}}].

\bibitem{Balaguera_2020}
A.~{Balaguera-Antol{\'\i}nez}, F.-S.~{Kitaura}, M.~{Pellejero-Ib{\'a}{\~n}ez},
  M.~{Lippich}, C.~{Zhao}, A.G.~{S{\'a}nchez} et~al., \emph{{One simulation to
  have them all: performance of the Bias Assignment Method against N-body
  simulations}}, \href{https://doi.org/10.1093/mnras/stz3206}{\emph{\mnras}
  {\bfseries 491} (2020) 2565}
  [\href{https://arxiv.org/abs/1906.06109}{{\ttfamily 1906.06109}}].

\bibitem{Kitaura_2022}
F.S.~{Kitaura}, A.~{Balaguera-Antol{\'\i}nez}, F.~{Sinigaglia} and
  M.~{Pellejero-Ib{\'a}{\~n}ez}, \emph{{The cosmic web connection to the dark
  matter halo distribution through gravity}},
  \href{https://doi.org/10.1093/mnras/stac671}{\emph{\mnras} {\bfseries 512}
  (2022) 2245} [\href{https://arxiv.org/abs/2005.11598}{{\ttfamily
  2005.11598}}].

\bibitem{Balaguera_2023}
A.~{Balaguera-Antol{\'\i}nez}, F.-S.~{Kitaura}, S.~{Alam}, C.-H.~{Chuang},
  Y.~{Yu}, G.~{Favole} et~al., \emph{{DESI mock challenge. Halo and galaxy
  catalogues with the bias assignment method}},
  \href{https://doi.org/10.1051/0004-6361/202245618}{\emph{\aap} {\bfseries
  673} (2023) A130} [\href{https://arxiv.org/abs/2211.10640}{{\ttfamily
  2211.10640}}].

\bibitem{Chuang_2015}
C.-H.~{Chuang}, C.~{Zhao}, F.~{Prada}, E.~{Munari}, S.~{Avila}, A.~{Izard}
  et~al., \emph{{nIFTy cosmology: Galaxy/halo mock catalogue comparison project
  on clustering statistics}},
  \href{https://doi.org/10.1093/mnras/stv1289}{\emph{\mnras} {\bfseries 452}
  (2015) 686} [\href{https://arxiv.org/abs/1412.7729}{{\ttfamily 1412.7729}}].

\bibitem{Lippich_2019}
M.~{Lippich}, A.G.~{S{\'a}nchez}, M.~{Colavincenzo}, E.~{Sefusatti},
  P.~{Monaco}, L.~{Blot} et~al., \emph{{Comparing approximate methods for mock
  catalogues and covariance matrices - I. Correlation function}},
  \href{https://doi.org/10.1093/mnras/sty2757}{\emph{\mnras} {\bfseries 482}
  (2019) 1786} [\href{https://arxiv.org/abs/1806.09477}{{\ttfamily
  1806.09477}}].

\bibitem{Blot_2019}
L.~{Blot}, M.~{Crocce}, E.~{Sefusatti}, M.~{Lippich}, A.G.~{S{\'a}nchez},
  M.~{Colavincenzo} et~al., \emph{{Comparing approximate methods for mock
  catalogues and covariance matrices II: power spectrum multipoles}},
  \href{https://doi.org/10.1093/mnras/stz507}{\emph{\mnras} {\bfseries 485}
  (2019) 2806} [\href{https://arxiv.org/abs/1806.09497}{{\ttfamily
  1806.09497}}].

\bibitem{Colavincenzo_2019}
M.~{Colavincenzo}, E.~{Sefusatti}, P.~{Monaco}, L.~{Blot}, M.~{Crocce},
  M.~{Lippich} et~al., \emph{{Comparing approximate methods for mock catalogues
  and covariance matrices - III: bispectrum}},
  \href{https://doi.org/10.1093/mnras/sty2964}{\emph{\mnras} {\bfseries 482}
  (2019) 4883} [\href{https://arxiv.org/abs/1806.09499}{{\ttfamily
  1806.09499}}].

\bibitem{Kitaura_2021}
F.-S.~{Kitaura}, M.~{Ata}, S.A.~{Rodr{\'\i}guez-Torres},
  M.~{Hern{\'a}ndez-S{\'a}nchez}, A.~{Balaguera-Antol{\'\i}nez} and G.~{Yepes},
  \emph{{COSMIC BIRTH: efficient Bayesian inference of the evolving cosmic web
  from galaxy surveys}},
  \href{https://doi.org/10.1093/mnras/staa3774}{\emph{\mnras} {\bfseries 502}
  (2021) 3456} [\href{https://arxiv.org/abs/1911.00284}{{\ttfamily
  1911.00284}}].

\bibitem{Ata_2015}
M.~{Ata}, F.-S.~{Kitaura} and V.~{M{\"u}ller}, \emph{{Bayesian inference of
  cosmic density fields from non-linear, scale-dependent, and stochastic biased
  tracers}}, \href{https://doi.org/10.1093/mnras/stu2347}{\emph{\mnras}
  {\bfseries 446} (2015) 4250}
  [\href{https://arxiv.org/abs/1408.2566}{{\ttfamily 1408.2566}}].

\bibitem{Kitaura_2013}
F.-S.~Kitaura and S.~He{\ss}, \emph{Cosmological structure formation with
  augmented lagrangian perturbation theory},
  \href{https://doi.org/10.1093/mnrasl/slt101}{\emph{Monthly Notices of the
  Royal Astronomical Society: Letters} {\bfseries 435} (2013) L78}.

\bibitem{Kitaura_2024}
F.S.~{Kitaura}, F.~{Sinigaglia}, A.~{Balaguera-Antol{\'\i}nez} and G.~{Favole},
  \emph{{The cosmic web from perturbation theory}},
  \href{https://doi.org/10.1051/0004-6361/202345876}{\emph{\aap} {\bfseries
  683} (2024) A215} [\href{https://arxiv.org/abs/2301.03648}{{\ttfamily
  2301.03648}}].

\bibitem{Feng_2016}
Y.~{Feng}, M.-Y.~{Chu}, U.~{Seljak} and P.~{McDonald}, \emph{{FASTPM: a new
  scheme for fast simulations of dark matter and haloes}},
  \href{https://doi.org/10.1093/mnras/stw2123}{\emph{Monthly Notices of the
  Royal Astronomical Society} {\bfseries 463} (2016) 2273}
  [\href{https://arxiv.org/abs/1603.00476}{{\ttfamily 1603.00476}}].

\bibitem{Tassev_2013}
S.~{Tassev}, M.~{Zaldarriaga} and D.J.~{Eisenstein}, \emph{{Solving large scale
  structure in ten easy steps with COLA}},
  \href{https://doi.org/10.1088/1475-7516/2013/06/036}{\emph{Journal of Cosmoly
  and Astroparticle Physics} {\bfseries 6} (2013) 036}
  [\href{https://arxiv.org/abs/1301.0322}{{\ttfamily 1301.0322}}].

\bibitem{Pellejero_2020}
M.~Pellejero-Ibañez, A.~Balaguera-Antolínez, F.-S.~Kitaura, R.E.~Angulo,
  G.~Yepes, C.-H.~Chuang et~al., \emph{{The bias of dark matter tracers:
  assessing the accuracy of mapping techniques}},
  \href{https://doi.org/10.1093/mnras/staa270}{\emph{Monthly Notices of the
  Royal Astronomical Society} {\bfseries 493} (2020) 586}
  [\href{https://arxiv.org/abs/https://academic.oup.com/mnras/article-pdf/493/1/586/32513186/staa270.pdf}{{\ttfamily
  https://academic.oup.com/mnras/article-pdf/493/1/586/32513186/staa270.pdf}}].

\bibitem{Sinigaglia_2021}
F.~Sinigaglia, F.-S.~Kitaura, A.~Balaguera-Antolínez, K.~Nagamine, M.~Ata,
  I.~Shimizu et~al., \emph{The bias from hydrodynamic simulations: Mapping
  baryon physics onto dark matter fields},
  \href{https://doi.org/10.3847/1538-4357/ac158b}{\emph{The Astrophysical
  Journal} {\bfseries 921} (2021) 66}.

\bibitem{Sinigaglia_2022}
F.~{Sinigaglia}, F.-S.~{Kitaura}, A.~{Balaguera-Antol{\'\i}nez}, I.~{Shimizu},
  K.~{Nagamine}, M.~{S{\'a}nchez-Benavente} et~al., \emph{{Mapping the
  Three-dimensional Ly{\ensuremath{\alpha}} Forest Large-scale Structure in
  Real and Redshift Space}},
  \href{https://doi.org/10.3847/1538-4357/ac5112}{\emph{\apj} {\bfseries 927}
  (2022) 230} [\href{https://arxiv.org/abs/2107.07917}{{\ttfamily
  2107.07917}}].

\bibitem{Sinigaglia_2024}
F.~{Sinigaglia}, F.S.~{Kitaura}, K.~{Nagamine}, Y.~{Oku} and
  A.~{Balaguera-Antol{\'\i}nez}, \emph{{Field-level Lyman-{\ensuremath{\alpha}}
  forest modeling in redshift space via augmented nonlocal Fluctuating
  Gunn-Peterson Approximation}},
  \href{https://doi.org/10.1051/0004-6361/202346931}{\emph{\aap} {\bfseries
  682} (2024) A21} [\href{https://arxiv.org/abs/2305.10428}{{\ttfamily
  2305.10428}}].

\bibitem{Chuang_2019MNRAS}
C.-H.~{Chuang}, G.~{Yepes}, F.-S.~{Kitaura}, M.~{Pellejero-Ibanez},
  S.~{Rodr{\'\i}guez-Torres}, Y.~{Feng} et~al., \emph{{UNIT project: Universe
  N-body simulations for the Investigation of Theoretical models from galaxy
  surveys}}, \href{https://doi.org/10.1093/mnras/stz1233}{\emph{\mnras}
  {\bfseries 487} (2019) 48}
  [\href{https://arxiv.org/abs/1811.02111}{{\ttfamily 1811.02111}}].

\bibitem{McDonald_2009}
P.~McDonald and A.~Roy, \emph{Clustering of dark matter tracers: generalizing
  bias for the coming era of precision {LSS}},
  \href{https://doi.org/10.1088/1475-7516/2009/08/020}{\emph{Journal of
  Cosmology and Astroparticle Physics} {\bfseries 2009} (2009) 020}.

\bibitem{2006PhRvD..73f3519C}
M.~{Crocce} and R.~{Scoccimarro}, \emph{{Renormalized cosmological perturbation
  theory}}, \href{https://doi.org/10.1103/PhysRevD.73.063519}{\emph{\prd}
  {\bfseries 73} (2006) 063519}
  [\href{https://arxiv.org/abs/astro-ph/0509418}{{\ttfamily
  astro-ph/0509418}}].

\bibitem{Baumann_2012}
D.~{Baumann}, A.~{Nicolis}, L.~{Senatore} and M.~{Zaldarriaga},
  \emph{{Cosmological non-linearities as an effective fluid}},
  \href{https://doi.org/10.1088/1475-7516/2012/07/051}{\emph{\jcap} {\bfseries
  2012} (2012) 051} [\href{https://arxiv.org/abs/1004.2488}{{\ttfamily
  1004.2488}}].

\bibitem{Carrasco_2012}
J.J.M.~{Carrasco}, M.P.~{Hertzberg} and L.~{Senatore}, \emph{{The effective
  field theory of cosmological large scale structures}},
  \href{https://doi.org/10.1007/JHEP09(2012)082}{\emph{Journal of High Energy
  Physics} {\bfseries 2012} (2012) 82}
  [\href{https://arxiv.org/abs/1206.2926}{{\ttfamily 1206.2926}}].

\bibitem{Pajer_2013}
E.~{Pajer} and M.~{Zaldarriaga}, \emph{{On the renormalization of the effective
  field theory of large scale structures}},
  \href{https://doi.org/10.1088/1475-7516/2013/08/037}{\emph{\jcap} {\bfseries
  2013} (2013) 037} [\href{https://arxiv.org/abs/1301.7182}{{\ttfamily
  1301.7182}}].

\bibitem{Porto_2014}
R.A.~{Porto}, L.~{Senatore} and M.~{Zaldarriaga}, \emph{{The Lagrangian-space
  Effective Field Theory of large scale structures}},
  \href{https://doi.org/10.1088/1475-7516/2014/05/022}{\emph{\jcap} {\bfseries
  2014} (2014) 022} [\href{https://arxiv.org/abs/1311.2168}{{\ttfamily
  1311.2168}}].

\bibitem{Angulo_2015}
R.~{Angulo}, M.~{Fasiello}, L.~{Senatore} and Z.~{Vlah}, \emph{{On the
  statistics of biased tracers in the Effective Field Theory of Large Scale
  Structures}},
  \href{https://doi.org/10.1088/1475-7516/2015/09/029}{\emph{\jcap} {\bfseries
  2015} (2015) 029} [\href{https://arxiv.org/abs/1503.08826}{{\ttfamily
  1503.08826}}].

\bibitem{Senatore_2015}
L.~{Senatore} and M.~{Zaldarriaga}, \emph{{The IR-resummed Effective Field
  Theory of Large Scale Structures}},
  \href{https://doi.org/10.1088/1475-7516/2015/02/013}{\emph{\jcap} {\bfseries
  2015} (2015) 013} [\href{https://arxiv.org/abs/1404.5954}{{\ttfamily
  1404.5954}}].

\bibitem{Senatore_2015b}
L.~{Senatore}, \emph{{Bias in the effective field theory of large scale
  structures}},
  \href{https://doi.org/10.1088/1475-7516/2015/11/007}{\emph{\jcap} {\bfseries
  2015} (2015) 007} [\href{https://arxiv.org/abs/1406.7843}{{\ttfamily
  1406.7843}}].

\bibitem{Vlah_2015}
Z.~{Vlah}, M.~{White} and A.~{Aviles}, \emph{{A Lagrangian effective field
  theory}}, \href{https://doi.org/10.1088/1475-7516/2015/09/014}{\emph{\jcap}
  {\bfseries 2015} (2015) 014}
  [\href{https://arxiv.org/abs/1506.05264}{{\ttfamily 1506.05264}}].

\bibitem{Schmittfull_2019}
M.~{Schmittfull}, M.~{Simonovi{\'c}}, V.~{Assassi} and M.~{Zaldarriaga},
  \emph{{Modeling biased tracers at the field level}},
  \href{https://doi.org/10.1103/PhysRevD.100.043514}{\emph{\prd} {\bfseries
  100} (2019) 043514} [\href{https://arxiv.org/abs/1811.10640}{{\ttfamily
  1811.10640}}].

\bibitem{Forero_2024}
D.~{Forero-S{\'a}nchez}, F.-S.~{Kitaura}, F.~{Sinigaglia}, J.~{Mar{\'\i}a
  Coloma-Nodal} and J.-P.~{Kneib}, \emph{{CosmoMIA: Cosmic Web-based redshift
  space halo distribution}},
  \href{https://doi.org/10.48550/arXiv.2402.17581}{\emph{arXiv e-prints} (2024)
  arXiv:2402.17581} [\href{https://arxiv.org/abs/2402.17581}{{\ttfamily
  2402.17581}}].

\bibitem{Buchert_1994}
T.~{Buchert}, \emph{{Lagrangian Theory of Gravitational Instability of
  Friedman-Lemaitre Cosmologies - a Generic Third-Order Model for Nonlinear
  Clustering}}, \href{https://doi.org/10.1093/mnras/267.4.811}{\emph{\mnras}
  {\bfseries 267} (1994) 811}
  [\href{https://arxiv.org/abs/astro-ph/9309055}{{\ttfamily
  astro-ph/9309055}}].

\bibitem{Bouchet_1995}
F.R.~{Bouchet}, S.~{Colombi}, E.~{Hivon} and R.~{Juszkiewicz},
  \emph{{Perturbative Lagrangian approach to gravitational instability.}},
  \href{https://doi.org/10.48550/arXiv.astro-ph/9406013}{\emph{\aap} {\bfseries
  296} (1995) 575} [\href{https://arxiv.org/abs/astro-ph/9406013}{{\ttfamily
  astro-ph/9406013}}].

\bibitem{Catelan_1995}
P.~{Catelan}, \emph{{Lagrangian dynamics in non-flat universes and non-linear
  gravitational evolution}},
  \href{https://doi.org/10.1093/mnras/276.1.115}{\emph{\mnras} {\bfseries 276}
  (1995) 115} [\href{https://arxiv.org/abs/astro-ph/9406016}{{\ttfamily
  astro-ph/9406016}}].

\bibitem{Crocce_2006}
M.~{Crocce}, S.~{Pueblas} and R.~{Scoccimarro}, \emph{{Transients from initial
  conditions in cosmological simulations}},
  \href{https://doi.org/10.1111/j.1365-2966.2006.11040.x}{\emph{\mnras}
  {\bfseries 373} (2006) 369}
  [\href{https://arxiv.org/abs/astro-ph/0606505}{{\ttfamily
  astro-ph/0606505}}].

\bibitem{Bernardeau_1994}
F.~{Bernardeau}, \emph{{The Nonlinear Evolution of Rare Events}},
  \href{https://doi.org/10.1086/174121}{\emph{\apj} {\bfseries 427} (1994) 51}
  [\href{https://arxiv.org/abs/astro-ph/9311066}{{\ttfamily
  astro-ph/9311066}}].

\bibitem{Mohayaee_2006}
R.~Mohayaee, H.~Mathis, S.~Colombi and J.~Silk, \emph{{Reconstruction of
  primordial density fields}},
  \href{https://doi.org/10.1111/j.1365-2966.2005.09774.x}{\emph{Monthly Notices
  of the Royal Astronomical Society} {\bfseries 365} (2006) 939}
  [\href{https://arxiv.org/abs/https://academic.oup.com/mnras/article-pdf/365/3/939/2910100/365-3-939.pdf}{{\ttfamily
  https://academic.oup.com/mnras/article-pdf/365/3/939/2910100/365-3-939.pdf}}].

\bibitem{neyrinck_2013}
M.C.~Neyrinck, \emph{Quantifying distortions of the lagrangian dark-matter mesh
  in cosmology}, {\emph{Monthly Notices of the Royal Astronomical Society}
  {\bfseries 428} (2013) 141}.

\bibitem{Abel_2012}
T.~{Abel}, O.~{Hahn} and R.~{Kaehler}, \emph{{Tracing the dark matter sheet in
  phase space}},
  \href{https://doi.org/10.1111/j.1365-2966.2012.21754.x}{\emph{Monthly Notices
  of the Royal Astronomical Society} {\bfseries 427} (2012) 61}
  [\href{https://arxiv.org/abs/1111.3944}{{\ttfamily 1111.3944}}].

\bibitem{Neyrinck_2016}
M.C.~{Neyrinck}, \emph{{Truthing the stretch: non-perturbative cosmological
  realizations with multiscale spherical collapse}},
  \href{https://doi.org/10.1093/mnrasl/slv141}{\emph{\mnras} {\bfseries 455}
  (2016) L11} [\href{https://arxiv.org/abs/1503.07534}{{\ttfamily
  1503.07534}}].

\bibitem{Minerva_2016}
J.N.~{Grieb}, A.G.~{S{\'a}nchez}, S.~{Salazar-Albornoz} and C.~{Dalla Vecchia},
  \emph{{Gaussian covariance matrices for anisotropic galaxy clustering
  measurements}}, \href{https://doi.org/10.1093/mnras/stw065}{\emph{\mnras}
  {\bfseries 457} (2016) 1577}
  [\href{https://arxiv.org/abs/1509.04293}{{\ttfamily 1509.04293}}].

\bibitem{Desjacques_2018}
V.~{Desjacques}, D.~{Jeong} and F.~{Schmidt}, \emph{{Large-scale galaxy bias}},
  \href{https://doi.org/10.1016/j.physrep.2017.12.002}{\emph{\physrep}
  {\bfseries 733} (2018) 1} [\href{https://arxiv.org/abs/1611.09787}{{\ttfamily
  1611.09787}}].

\bibitem{Gao_2005}
L.~{Gao}, V.~{Springel} and S.D.M.~{White}, \emph{{The age dependence of halo
  clustering}},
  \href{https://doi.org/10.1111/j.1745-3933.2005.00084.x}{\emph{\mnras}
  {\bfseries 363} (2005) L66}
  [\href{https://arxiv.org/abs/astro-ph/0506510}{{\ttfamily
  astro-ph/0506510}}].

\bibitem{Croton_2007}
D.J.~{Croton}, L.~{Gao} and S.D.M.~{White}, \emph{{Halo assembly bias and its
  effects on galaxy clustering}},
  \href{https://doi.org/10.1111/j.1365-2966.2006.11230.x}{\emph{\mnras}
  {\bfseries 374} (2007) 1303}
  [\href{https://arxiv.org/abs/astro-ph/0605636}{{\ttfamily
  astro-ph/0605636}}].

\bibitem{Dalal_2008}
N.~{Dalal}, M.~{White}, J.R.~{Bond} and A.~{Shirokov}, \emph{{Halo Assembly
  Bias in Hierarchical Structure Formation}},
  \href{https://doi.org/10.1086/591512}{\emph{\apj} {\bfseries 687} (2008) 12}
  [\href{https://arxiv.org/abs/0803.3453}{{\ttfamily 0803.3453}}].

\bibitem{Tojeiro_2017}
R.~{Tojeiro}, E.~{Eardley}, J.A.~{Peacock}, P.~{Norberg}, M.~{Alpaslan},
  S.P.~{Driver} et~al., \emph{{Galaxy and Mass Assembly (GAMA): halo formation
  times and halo assembly bias on the cosmic web}},
  \href{https://doi.org/10.1093/mnras/stx1466}{\emph{\mnras} {\bfseries 470}
  (2017) 3720} [\href{https://arxiv.org/abs/1612.08595}{{\ttfamily
  1612.08595}}].

\bibitem{Yang_2017}
X.~Yang, Y.~Zhang, T.~Lu, H.~Wang, F.~Shi, D.~Tweed et~al., \emph{Revealing the
  cosmic web-dependent halo bias},
  \href{https://doi.org/10.3847/1538-4357/aa8c7a}{\emph{The Astrophysical
  Journal} {\bfseries 848} (2017) 60}.

\bibitem{Fry_1993}
J.N.~{Fry} and E.~{Gaztanaga}, \emph{{Biasing and Hierarchical Statistics in
  Large-Scale Structure}}, \href{https://doi.org/10.1086/173015}{\emph{\apj}
  {\bfseries 413} (1993) 447}
  [\href{https://arxiv.org/abs/astro-ph/9302009}{{\ttfamily
  astro-ph/9302009}}].

\bibitem{Cen_1993}
R.~{Cen} and J.P.~{Ostriker}, \emph{{Cold Dark Matter Cosmogony with
  Hydrodynamics and Galaxy Formation: Galaxy Properties at Redshift Zero}},
  \href{https://doi.org/10.1086/173322}{\emph{\apj} {\bfseries 417} (1993)
  415}.

\bibitem{delatorre_2013}
S.~{de la Torre} and J.A.~{Peacock}, \emph{{Reconstructing the distribution of
  haloes and mock galaxies below the resolution limit in cosmological
  simulations}}, \href{https://doi.org/10.1093/mnras/stt1333}{\emph{\mnras}
  {\bfseries 435} (2013) 743}
  [\href{https://arxiv.org/abs/1212.3615}{{\ttfamily 1212.3615}}].

\bibitem{Kaiser_1984}
N.~{Kaiser}, \emph{{On the spatial correlations of Abell clusters.}},
  \href{https://doi.org/10.1086/184341}{\emph{\apjl} {\bfseries 284} (1984)
  L9}.

\bibitem{Neyrinck_2014}
M.C.~{Neyrinck}, M.A.~{Arag{\'o}n-Calvo}, D.~{Jeong} and X.~{Wang}, \emph{{A
  halo bias function measured deeply into voids without stochasticity}},
  \href{https://doi.org/10.1093/mnras/stu589}{\emph{\mnras} {\bfseries 441}
  (2014) 646} [\href{https://arxiv.org/abs/1309.6641}{{\ttfamily 1309.6641}}].

\bibitem{Garcia_2019}
R.~{Garc{\'\i}a} and E.~{Rozo}, \emph{{Halo exclusion criteria impacts halo
  statistics}}, \href{https://doi.org/10.1093/mnras/stz2458}{\emph{\mnras}
  {\bfseries 489} (2019) 4170}
  [\href{https://arxiv.org/abs/1903.01709}{{\ttfamily 1903.01709}}].

\bibitem{Dekel_1999}
A.~{Dekel} and O.~{Lahav}, \emph{{Stochastic Nonlinear Galaxy Biasing}},
  \href{https://doi.org/10.1086/307428}{\emph{\apj} {\bfseries 520} (1999) 24}
  [\href{https://arxiv.org/abs/astro-ph/9806193}{{\ttfamily
  astro-ph/9806193}}].

\bibitem{Sheth_1999}
R.K.~{Sheth} and G.~{Lemson}, \emph{{Biasing and the distribution of dark
  matter haloes}},
  \href{https://doi.org/10.1046/j.1365-8711.1999.02378.x}{\emph{\mnras}
  {\bfseries 304} (1999) 767}
  [\href{https://arxiv.org/abs/astro-ph/9808138}{{\ttfamily
  astro-ph/9808138}}].

\bibitem{Peebles_1980}
P.J.E.~{Peebles}, \emph{{The large-scale structure of the universe}}, Princeton
  university press (1980).

\bibitem{Somerville_2001}
R.S.~{Somerville}, G.~{Lemson}, Y.~{Sigad}, A.~{Dekel}, G.~{Kauffmann} and
  S.D.M.~{White}, \emph{{Non-linear stochastic galaxy biasing in cosmological
  simulations}},
  \href{https://doi.org/10.1046/j.1365-8711.2001.03894.x}{\emph{\mnras}
  {\bfseries 320} (2001) 289}
  [\href{https://arxiv.org/abs/astro-ph/9912073}{{\ttfamily
  astro-ph/9912073}}].

\bibitem{Casas_Miranda_2002}
R.~Casas-Miranda, H.J.~Mo, R.K.~Sheth and G.~Boerner, \emph{On the distribution
  of haloes, galaxies and mass},
  \href{https://doi.org/10.1046/j.1365-8711.2002.05378.x}{\emph{Monthly Notices
  of the Royal Astronomical Society} {\bfseries 333} (2002) 730–738}.

\bibitem{Saslaw_1984}
W.C.~{Saslaw} and A.J.S.~{Hamilton}, \emph{{Thermodynamics and galaxy
  clustering - Nonlinear theory of high order correlations}},
  \href{https://doi.org/10.1086/161589}{\emph{\apj} {\bfseries 276} (1984) 13}.

\bibitem{Sheth_1995}
R.K.~Sheth, \emph{{Press–Schechter, thermodynamics and gravitational
  clustering}}, \href{https://doi.org/10.1093/mnras/274.1.213}{\emph{Monthly
  Notices of the Royal Astronomical Society} {\bfseries 274} (1995) 213}
  [\href{https://arxiv.org/abs/https://academic.oup.com/mnras/article-pdf/274/1/213/18539982/mnras274-0213.pdf}{{\ttfamily
  https://academic.oup.com/mnras/article-pdf/274/1/213/18539982/mnras274-0213.pdf}}].

\bibitem{wild_2005}
V.~Wild, J.A.~Peacock, O.~Lahav, E.~Conway, S.~Maddox, I.K.~Baldry et~al.,
  \emph{{The 2dF Galaxy Redshift Survey: stochastic relative biasing between
  galaxy populations}},
  \href{https://doi.org/10.1111/j.1365-2966.2004.08447.x}{\emph{Monthly Notices
  of the Royal Astronomical Society} {\bfseries 356} (2005) 247}
  [\href{https://arxiv.org/abs/https://academic.oup.com/mnras/article-pdf/356/1/247/3589649/356-1-247.pdf}{{\ttfamily
  https://academic.oup.com/mnras/article-pdf/356/1/247/3589649/356-1-247.pdf}}].

\bibitem{de_la_Torre_2013}
S.~de~la Torre and J.A.~Peacock, \emph{Reconstructing the distribution of
  haloes and mock galaxies below the resolution limit in cosmological
  simulations}, \href{https://doi.org/10.1093/mnras/stt1333}{\emph{Monthly
  Notices of the Royal Astronomical Society} {\bfseries 435} (2013) 743}.

\bibitem{Bond_1996nat}
J.R.~{Bond}, L.~{Kofman} and D.~{Pogosyan}, \emph{{How filaments of galaxies
  are woven into the cosmic web}},
  \href{https://doi.org/10.1038/380603a0}{\emph{\nat} {\bfseries 380} (1996)
  603} [\href{https://arxiv.org/abs/astro-ph/9512141}{{\ttfamily
  astro-ph/9512141}}].

\bibitem{van_de_Weygaert_2008}
R.~van~de Weygaert and W.~Schaap, \emph{The cosmic web: Geometric analysis},
  in \emph{Data Analysis in Cosmology}, pp.~291--413, Springer Berlin
  Heidelberg (2008), \href{https://doi.org/10.1007/978-3-540-44767-2_11}{DOI}.

\bibitem{Gott_2005}
I.~{Gott}, J.~Richard, M.~{Juri{\'c}}, D.~{Schlegel}, F.~{Hoyle}, M.~{Vogeley},
  M.~{Tegmark} et~al., \emph{{A Map of the Universe}},
  \href{https://doi.org/10.1086/428890}{\emph{\apj} {\bfseries 624} (2005) 463}
  [\href{https://arxiv.org/abs/astro-ph/0310571}{{\ttfamily
  astro-ph/0310571}}].

\bibitem{Zeldovich_1970}
Y.B.~{Zel'dovich}, \emph{{Gravitational instability: An approximate theory for
  large density perturbations.}}, {\emph{\aap} {\bfseries 5} (1970) 84}.

\bibitem{Hahn_2007}
O.~Hahn, C.~Porciani, C.M.~Carollo and A.~Dekel, \emph{Properties of dark
  matter haloes in clusters, filaments, sheets and voids},
  \href{https://doi.org/10.1111/j.1365-2966.2006.11318.x}{\emph{Monthly Notices
  of the Royal Astronomical Society} {\bfseries 375} (2007) 489}.

\bibitem{Forero_2009}
J.E.~Forero–Romero, Y.~Hoffman, S.~Gottlöber, A.~Klypin and G.~Yepes,
  \emph{{A dynamical classification of the cosmic web}},
  \href{https://doi.org/10.1111/j.1365-2966.2009.14885.x}{\emph{Monthly Notices
  of the Royal Astronomical Society} {\bfseries 396} (2009) 1815}
  [\href{https://arxiv.org/abs/https://academic.oup.com/mnras/article-pdf/396/3/1815/5804803/mnras0396-1815.pdf}{{\ttfamily
  https://academic.oup.com/mnras/article-pdf/396/3/1815/5804803/mnras0396-1815.pdf}}].

\bibitem{Martizzi_2019}
D.~Martizzi, M.~Vogelsberger, M.C.~Artale, M.~Haider, P.~Torrey, F.~Marinacci
  et~al., \emph{{Baryons in the Cosmic Web of IllustrisTNG – I: gas in knots,
  filaments, sheets, and voids}},
  \href{https://doi.org/10.1093/mnras/stz1106}{\emph{Monthly Notices of the
  Royal Astronomical Society} {\bfseries 486} (2019) 3766}
  [\href{https://arxiv.org/abs/https://academic.oup.com/mnras/article-pdf/486/3/3766/28548252/stz1106.pdf}{{\ttfamily
  https://academic.oup.com/mnras/article-pdf/486/3/3766/28548252/stz1106.pdf}}].

\bibitem{Zhao_2015}
C.~Zhao, F.-S.~Kitaura, C.-H.~Chuang, F.~Prada, G.~Yepes and C.~Tao,
  \emph{{Halo mass distribution reconstruction across the cosmic web}},
  \href{https://doi.org/10.1093/mnras/stv1262}{\emph{Monthly Notices of the
  Royal Astronomical Society} {\bfseries 451} (2015) 4266}
  [\href{https://arxiv.org/abs/https://academic.oup.com/mnras/article-pdf/451/4/4266/3901283/stv1262.pdf}{{\ttfamily
  https://academic.oup.com/mnras/article-pdf/451/4/4266/3901283/stv1262.pdf}}].

\bibitem{Lee_2016}
K.-G.~{Lee} and M.~{White}, \emph{{Revealing the z
  \raisebox{-0.5ex}\textasciitilde 2.5 Cosmic Web with 3D
  Ly{\ensuremath{\alpha}} Forest Tomography: a Deformation Tensor Approach}},
  \href{https://doi.org/10.3847/0004-637X/831/2/181}{\emph{\apj} {\bfseries
  831} (2016) 181} [\href{https://arxiv.org/abs/1603.04441}{{\ttfamily
  1603.04441}}].

\bibitem{Krolewski_2017}
A.~{Krolewski}, K.-G.~{Lee}, Z.~{Luki{\'c}} and M.~{White}, \emph{{Measuring
  Alignments between Galaxies and the Cosmic Web at z {\ensuremath{\sim}} 2-3
  Using IGM Tomography}},
  \href{https://doi.org/10.3847/1538-4357/837/1/31}{\emph{\apj} {\bfseries 837}
  (2017) 31} [\href{https://arxiv.org/abs/1612.00067}{{\ttfamily 1612.00067}}].

\bibitem{Horowitz_2019}
B.~{Horowitz}, K.-G.~{Lee}, M.~{White}, A.~{Krolewski} and M.~{Ata},
  \emph{{TARDIS. I. A Constrained Reconstruction Approach to Modeling the z
  {\ensuremath{\sim}} 2.5 Cosmic Web Probed by Ly{\ensuremath{\alpha}} Forest
  Tomography}}, \href{https://doi.org/10.3847/1538-4357/ab4d4c}{\emph{\apj}
  {\bfseries 887} (2019) 61}
  [\href{https://arxiv.org/abs/1903.09049}{{\ttfamily 1903.09049}}].

\bibitem{Planck_2016A}
{Planck Collaboration}, P.A.R.~{Ade}, N.~{Aghanim}, M.~{Arnaud}, M.~{Ashdown},
  J.~{Aumont} et~al., \emph{{Planck 2015 results. XIII. Cosmological
  parameters}}, \href{https://doi.org/10.1051/0004-6361/201525830}{\emph{\aap}
  {\bfseries 594} (2016) A13}
  [\href{https://arxiv.org/abs/1502.01589}{{\ttfamily 1502.01589}}].

\bibitem{Feng_2016MNRAS}
Y.~{Feng}, M.-Y.~{Chu}, U.~{Seljak} and P.~{McDonald}, \emph{{FASTPM: a new
  scheme for fast simulations of dark matter and haloes}},
  \href{https://doi.org/10.1093/mnras/stw2123}{\emph{\mnras} {\bfseries 463}
  (2016) 2273} [\href{https://arxiv.org/abs/1603.00476}{{\ttfamily
  1603.00476}}].

\bibitem{Springel2005MNRAS}
V.~{Springel}, \emph{{The cosmological simulation code GADGET-2}},
  \href{https://doi.org/10.1111/j.1365-2966.2005.09655.x}{\emph{\mnras}
  {\bfseries 364} (2005) 1105}
  [\href{https://arxiv.org/abs/astro-ph/0505010}{{\ttfamily
  astro-ph/0505010}}].

\bibitem{Behroozi2013ApJ}
P.S.~{Behroozi}, R.H.~{Wechsler} and H.-Y.~{Wu}, \emph{{The ROCKSTAR
  Phase-space Temporal Halo Finder and the Velocity Offsets of Cluster Cores}},
  \href{https://doi.org/10.1088/0004-637X/762/2/109}{\emph{\apj} {\bfseries
  762} (2013) 109} [\href{https://arxiv.org/abs/1110.4372}{{\ttfamily
  1110.4372}}].

\bibitem{BehrooziConsist2013ApJ}
P.S.~{Behroozi}, R.H.~{Wechsler}, H.-Y.~{Wu}, M.T.~{Busha}, A.A.~{Klypin} and
  J.R.~{Primack}, \emph{{Gravitationally Consistent Halo Catalogs and Merger
  Trees for Precision Cosmology}},
  \href{https://doi.org/10.1088/0004-637X/763/1/18}{\emph{\apj} {\bfseries 763}
  (2013) 18} [\href{https://arxiv.org/abs/1110.4370}{{\ttfamily 1110.4370}}].

\bibitem{Alam_2020MNRAS}
S.~{Alam}, J.A.~{Peacock}, K.~{Kraljic}, A.J.~{Ross} and J.~{Comparat},
  \emph{{Multitracer extension of the halo model: probing quenching and
  conformity in eBOSS}},
  \href{https://doi.org/10.1093/mnras/staa1956}{\emph{\mnras} {\bfseries 497}
  (2020) 581} [\href{https://arxiv.org/abs/1910.05095}{{\ttfamily
  1910.05095}}].

\bibitem{KitauraSinigaglia_2024b}
{F.-S. Kitaura, F. Sinigaglia, et al.}{\emph{, in preparation} (2024) }.

\bibitem{Uhlemann_2020}
C.~{Uhlemann}, O.~{Friedrich}, F.~{Villaescusa-Navarro}, A.~{Banerjee} and
  S.~{Codis}, \emph{{Fisher for complements: extracting cosmology and neutrino
  mass from the counts-in-cells PDF}},
  \href{https://doi.org/10.1093/mnras/staa1155}{\emph{\mnras} {\bfseries 495}
  (2020) 4006} [\href{https://arxiv.org/abs/1911.11158}{{\ttfamily
  1911.11158}}].

\bibitem{Alam_2017}
S.~Alam, M.~Ata, S.~Bailey, F.~Beutler, D.~Bizyaev, J.A.~Blazek et~al.,
  \emph{The clustering of galaxies in the completed sdss-iii baryon oscillation
  spectroscopic survey: cosmological analysis of the dr12 galaxy sample},
  \href{https://doi.org/10.1093/mnras/stx721}{\emph{Monthly Notices of the
  Royal Astronomical Society} {\bfseries 470} (2017) 2617–2652}.

\bibitem{Sefusatti_2006}
E.~Sefusatti, M.~Crocce, S.~Pueblas and R.~Scoccimarro, \emph{Cosmology and the
  bispectrum}, \href{https://doi.org/10.1103/PhysRevD.74.023522}{\emph{Phys.
  Rev. D} {\bfseries 74} (2006) 023522}.

\bibitem{Sugiyama_2023}
N.S.~Sugiyama, D.~Yamauchi, T.~Kobayashi, T.~Fujita, S.~Arai, S.~Hirano et~al.,
  \emph{{New constraints on cosmological modified gravity theories from
  anisotropic three-point correlation functions of BOSS DR12 galaxies}},
  \href{https://doi.org/10.1093/mnras/stad1505}{\emph{Monthly Notices of the
  Royal Astronomical Society} {\bfseries 523} (2023) 3133}
  [\href{https://arxiv.org/abs/https://academic.oup.com/mnras/article-pdf/523/2/3133/50540934/stad1505.pdf}{{\ttfamily
  https://academic.oup.com/mnras/article-pdf/523/2/3133/50540934/stad1505.pdf}}].

\bibitem{Maksimova_2021}
N.A.~{Maksimova}, L.H.~{Garrison}, D.J.~{Eisenstein}, B.~{Hadzhiyska},
  S.~{Bose} and T.P.~{Satterthwaite}, \emph{{ABACUSSUMMIT: a massive set of
  high-accuracy, high-resolution N-body simulations}},
  \href{https://doi.org/10.1093/mnras/stab2484}{\emph{\mnras} {\bfseries 508}
  (2021) 4017} [\href{https://arxiv.org/abs/2110.11398}{{\ttfamily
  2110.11398}}].

\bibitem{GoodmanWeare_2010}
J.~{Goodman} and J.~{Weare}, \emph{{Ensemble samplers with affine invariance}},
  \href{https://doi.org/10.2140/camcos.2010.5.65}{\emph{Communications in
  Applied Mathematics and Computational Science} {\bfseries 5} (2010) 65}.

\bibitem{ForemanMackey_2013}
D.~{Foreman-Mackey}, D.W.~{Hogg}, D.~{Lang} and J.~{Goodman}, \emph{{emcee: The
  MCMC Hammer}}, \href{https://doi.org/10.1086/670067}{\emph{\pasp} {\bfseries
  125} (2013) 306} [\href{https://arxiv.org/abs/1202.3665}{{\ttfamily
  1202.3665}}].

\bibitem{Baumgarten_2018}
F.~{Baumgarten} and C.-H.~{Chuang}, \emph{{Robustness of the covariance matrix
  for galaxy clustering measurements}},
  \href{https://doi.org/10.1093/mnras/sty1971}{\emph{\mnras} {\bfseries 480}
  (2018) 2535} [\href{https://arxiv.org/abs/1802.04462}{{\ttfamily
  1802.04462}}].

\end{thebibliography}\endgroup

\end{document}